\documentclass[prb,twocolumn,superscriptaddress,showpacs]{revtex4-1}

\usepackage{graphicx}
\usepackage{amsmath,amssymb}
\usepackage{dcolumn}
\usepackage{bm}
\usepackage{color}
\usepackage{hyperref}
\hypersetup{colorlinks=true,breaklinks,linkcolor=blue,urlcolor=blue,citecolor=blue}
\usepackage{xspace}
\usepackage{multirow}

\DeclareRobustCommand{\vect}[1]{\bm{#1}}
\newcommand{\iw}{i\omega}
\newcommand{\iW}{i\Omega}

\newcommand\Xloc{X_\textrm{loc}}
\newcommand\Xq{X_{\bm{q}}}
\newcommand\bXloc{\bm{X}_\textrm{loc}}
\newcommand\bXlocZero{\bm{X}_\textrm{0,loc}}
\newcommand\bXq{\bm{X}_{\bm{q}}}
\newcommand\bXqZero{\bm{X}_{0,\bm{q}}}

\newcommand\chiloc{\chi_\textrm{loc}}
\newcommand\chiq{\chi_{\bm{q}}}
\newcommand\bchiloc{\bm{\chi}_\textrm{loc}}
\newcommand\bchiq{\bm{\chi}_{\bm{q}}}
\newcommand\chiQ{\chi_{\bm{Q}}}

\newcommand\Pq{\Lambda_{\bm{q}}}
\newcommand\Iq{I_{\bm{q}}}
\newcommand\IQ{I_{\bm{Q}}}

\newcommand\bPq{\bm{\Lambda}_{\bm{q}}}
\newcommand\bIq{\bm{I}_{\bm{q}}}

\newcommand\bphi{\bm{\phi}}
\newcommand\bPhi{\bm{\Phi}}

\newcommand\TN{\ensuremath{T_\textrm{N}}}

\begin{document}

\preprint{}

\title{Strong-coupling formula for momentum-dependent susceptibilities\\ in dynamical mean-field theory}

\author{Junya Otsuki}
\altaffiliation[Present address: ]{Research Institute for Interdisciplinary Science, Okayama University, Okayama 700-8530, Japan}
\affiliation{Department of Physics, Tohoku University, Sendai 980-8578, Japan}
\author{Kazuyoshi Yoshimi}
\affiliation{Institute for Solid State Physics, University of Tokyo, Chiba 277-8581, Japan}
\author{Hiroshi Shinaoka}
\affiliation{Department of Physics, Saitama University, Saitama 338-8570, Japan}
\author{Yusuke Nomura}
\affiliation{Department of Applied Physics, University of Tokyo, Hongo, Tokyo 113-8656, Japan}

\date{\today}

\begin{abstract}
Computing momentum-dependent susceptibilities in the dynamical mean-field theory (DMFT) requires solving the Bethe-Salpeter equation, which demands large computational cost.
Exploiting the strong-coupling feature of local fluctuations, we derive a simplified formula that can be solved at a considerably lower cost.
The validity and the physical meaning of the formula are confirmed by deriving the effective intersite interactions in the strong-coupling limit, such as the kinetic exchange and RKKY interactions.
Furthermore, numerical calculations for single-orbital and multiorbital models demonstrate surprisingly wider applicability including weak-coupling region.
Based on this formula, we propose three levels of practical approximations that can be chosen depending on complexity of problems.
Simpler evaluations of spin and orbital susceptibilities in multiorbital systems thus become possible within DMFT.
\end{abstract}

\maketitle

\section{Introduction}

Strongly correlated materials including $d$ and/or $f$ electrons show rich quantum phases that attract academic interest and promise novel functionality.
To find further interesting or useful properties, new compounds are searched for continuously.
On the theoretical side, a qualitative or even quantitative prediction of phase transitions is desired to promote experimental researches for finding new materials.

Dynamical mean-field theory (DMFT) is a method which is extensively employed in investigations of strongly correlated systems~\cite{Georges96}.
The DMFT takes full account of local correlations, which is responsible for formation of a local moment, and therefore offers a good starting point in addressing long-range ordering.
Applications of the DMFT have expanded from models to materials by combining first-principles calculations within the density functional theory (DFT+DMFT)\cite{Kotliar06}.

The DMFT is exploited not only for single-particle excitations but also for two-particle responses by means of the momentum-dependent susceptibilities $\chi(\bm{q},\omega)$.
The dynamical part gives information of two-particle excitations including collective modes.
On the other hand, the static component $\chi(\bm{q})$ allows systematic investigations of phase transitions.
This method has been applied to various correlated models such as the Hubbard model~\cite{Jarrell92,Lin12,Kunes14a,Geffroy19}, the periodic Anderson model~\cite{Jarrell95}, and variants of the Kondo lattice model~\cite{Otsuki09b,Otsuki09c,Hoshino10b,Hoshino13}.
It has also been applied to material calculations within the DFT+DMFT framework~\cite{Matsumoto09,Boehnke12,Yin14}.

An actual calculation of $\chi(\bm{q})$ requires much more effort compared to the ordinary self-consistency calculations:
In addition to the single-particle Green function (or the self-energy), one needs to compute the two-particle Green function (or the vertex part) which depends on three frequencies.
This is plugged into the Bethe-Salpeter (BS) equation to compute $\bm{q}$-dependent $\chi(\bm{q})$.
This additional process of computing $\chi(\bm{q})$ becomes particularly harder in multiorbital systems, since the number of susceptibility components increases in proportion to $N_\textrm{orb}^4$, where $N_\textrm{orb}$ is the number of orbitals.
Toward realistic susceptibility calculations including full orbitals, efforts have been made to develop an efficient treatment of the BS equation~\cite{Tagliavini18} and the vertex part~\cite{Kunes11a,Rohringer12,Kaufmann17,Wentzell-arXiv,Shinaoka18,Krien-arXiv}.
Related developments are addressed, for example, in solving the parquet equation~\cite{Li16} and diagrammatic extensions of the DMFT~\cite{Rohringer18,Galler18}.

In this paper, we establish a \emph{practical approximation} that enables quick computations of $\chi(\bm{q})$, rather than pursuing the rigorous solution of the original BS equation.
Using a strong-coupling feature of the vertex part, we first reduce the BS equation into a
simple form which is similar to the random phase approximation (RPA).
The difference with the RPA, however, is that the starting point of our formula is the local susceptibility, which fully takes local correlations into account.
The effective interaction becomes non-local, describing,
for example, the kinetic exchange interaction in the Hubbard model and the RKKY interaction in the periodic Anderson model. 
Thus, our simple formula captures correct physics of well-defined local moments in the strong-coupling limit (SCL).
For this reason, we refer to this formula as SCL formula.
It will be shown by numerical calculations that the SCL formula provides accurate results over wide range from weak- to strong-coupling region.

This paper is organized as follows.
In Sec.~\ref{sec:bse}, we first review necessary equations for computing $\chi(\bm{q})$, and present an explicit example that shows the issue to be addressed.
In Sec.~\ref{sec:scl}, we derive the SCL formula of $\chi(\bm{q})$
by reducing the BS equation.
In Sec.~\ref{sec:analytical}, we prove that the formula contains proper physics around the atomic limit such as the kinetic exchange interactions and the RKKY interactions.
The equation is extended to multiorbital systems in Sec.~\ref{sec:multiorbital}.
Based on the SCL
formula for $\chi(\bm{q})$, we propose three levels of practical approximations in Sec.~\ref{sec:approx}.
An application to a two-orbital model is presented in Sec.~\ref{sec:results-multiorb} to compare those approximation schemes.
The paper is summarized in Sec.~\ref{sec:summary}.

\section{Fundamentals}
\label{sec:bse}
\subsection{Susceptibility in the DMFT}

We first review how to evaluate the momentum dependent susceptibilities within the DMFT.
The basic idea is presented in Ref.~\onlinecite{Jarrell92}, which showed that the antiferromagnetic (AFM) transition temperature covers the paramagnetic Mott transition in the Hubbard model.
For early and recent reviews of the method, see Refs.~\onlinecite{Georges96, Kunes17}, respectively.

We define a local one-particle operator $O_{m\sigma, m'\sigma'}$ at site $i$ by
\begin{align}
O_{m\sigma,m'\sigma'}(i)=c_{im\sigma}^{\dag} c_{im'\sigma'},
\label{eq:O}
\end{align}
where $m$ and $\sigma$ denote orbital and spin components, respectively.
Any orbital or multipole operators that are invariant under the point-group symmetry are represented by a linear combinations of $O_{m\sigma,m'\sigma'}$.
In this paper, we consider momentum- and frequency-dependent susceptibility corresponding to $O_{m\sigma,m'\sigma'}$.
Regarding $O_{m\sigma,m'\sigma'}$ as a vector, the susceptibility is defined as a matrix $\bm{\chi}$ as
\begin{align}
[\bm{\chi}(\bm{q},\iW)]_{12,34}
= \int_0^{\beta} d\tau \langle O_{12}(\bm{q}, \tau) O_{43}(-\bm{q}) \rangle e^{\iW \tau},
\end{align}
where $\Omega$ denotes the bosonic Matsubara frequency, 
and abbreviations such as $1\equiv(m,\sigma)$ were introduced.
Eigenvalues of $\bm{\chi}$ gives physical susceptibilities $\chi^{(\gamma)}(\bm{q},\iW)$.
A phase transition can be detected by divergence of the static component $\chi^{(\gamma)}(\bm{q},0)$.

In the DMFT calculations, we treat the two-particle Green function, $X_{12,34}(\iw,\iw';\iW)$, defined with two additional fermionic Matsubara frequencies, $\iw$ and $\iw'$.
The physical susceptibility $\chi$ is obtained by summing over the fermionic frequencies as
\begin{align}
\chi_{12,34}(\bm{q},\iW)
= T\sum_{\omega\omega'} X_{12,34}(\iw, \iw'; \bm{q},\iW).
\label{eq:chi_and_X}
\end{align}
For convenience, we introduce a matrix notation including $\iw$ and $\iw'$ as
\begin{align}
X_{12,34}(\iw, \iw'; \bm{q},\iW) \equiv [\bm{X}(\bm{q},\iW)]_{(12,\omega), (34,\omega')}.
\end{align}
Then, the BS equation which $\bm{X}(\bm{q},\iW)$ satisfies is represented as a matrix equation of the form
\begin{align}
\bm{X}^{-1}(\bm{q}, \iW) &= \bm{X}_{0}^{-1}(\bm{q}, \iW) - \bm{\Gamma}_{\rm loc}(\iW),
\label{eq:BSE_lattice}
\end{align}
where $\bm{X}_{0}(\bm{q}, \iW)$ is defined by
\begin{align}
&[X_0(\bm{q}, \iW)]_{(12,\omega), (34,\omega')}
\nonumber \\
&= -\frac{\delta_{\omega \omega'}}{N} \sum_{\bm{k}} G_{31}(\bm{k}, \iw) G_{24}(\bm{k}+\bm{q}, \iw+\iW).
\end{align}
Here, $N$ denotes the number of lattice sites, and $G_{31}(\bm{k},\iw)$ is the single-particle Green function within DMFT.
The vertex part $\bm{\Gamma}_{\rm loc}(\iW)$ does not depend on momentum in accordance with the local self-energy in the DMFT.
Hence, $\bm{\Gamma}_{\rm loc}(\iW)$ can be evaluated in the effective impurity problem through a similar BS equation
\begin{align}
\bm{X}_{\textrm{loc}}^{-1}(\iW) &= \bm{X}_{0,\textrm{loc}}^{-1}(\iW) - \bm{\Gamma}_{\rm loc}(\iW),
\label{eq:BSE_loc}
\end{align}
where $\bXlocZero$ is defined by
$\bXlocZero(\iW) = N^{-1}\sum_{\bm{q}} \bm{X}_0(\bm{q}, \iW)$.
Eliminating $\bm{\Gamma}_{\rm loc}$ in Eqs.~(\ref{eq:BSE_lattice}) and (\ref{eq:BSE_loc}), we obtain
\begin{align}
\bm{X}^{-1}(\bm{q},\iW) &= \bm{X}_{\textrm{loc}}^{-1}(\iW) - \bm{X}_{0,\textrm{loc}}^{-1}(\iW) + \bm{X}_{0}^{-1}(\bm{q},\iW).
\label{eq:chiq-DMFT}
\end{align}
This is the equation we need to solve to compute $\bm{\chi}({\bm{q}})$.
The size of the matrices in Eq.~(\ref{eq:chiq-DMFT}) is, in general, $4 N_\textrm{orb}^2 N_{\omega} \times 4 N_\textrm{orb}^2 N_{\omega}$, where 4 comes from spin, $N_\textrm{orb}$ denotes the number of orbitals, and $N_{\omega}$ is the number of the fermionic Matsubara frequencies.
Solving the matrix equation as well as computing $\bXloc$ in the effective impurity problems becomes hard for multiorbital models at low temperatures.

Hereafter, we consider the static susceptibility, $\Omega=0$, and omit the argument $\iW$.

\subsection{Exemplary results and aim of this paper}
\label{sec:phase_rrpa}
To illustrate the formalism so far, and to make it clear the purpose of this paper,
we show exemplary results for the Hubbard model on a square lattice. The Hamiltonian reads
\begin{align}
\mathcal{H} = \sum_{\bm{k}\sigma} \epsilon_{\bm{k}} c_{\bm{k}\sigma}^{\dag} c_{\bm{k}\sigma} + U \sum_{i} n_{i\uparrow} n_{i\downarrow},
\label{eq:H-Hubbard}
\end{align}
where $\epsilon_{\bm{k}} = -2t(\cos k_x + \cos k_y)$.
The chemical potential is fixed at $\mu=U/2$ to consider a half-filled case.
The band width is given by $W=8t$, and we use $t=1$ as the unit of energy.
All computations were performed in a system of size $N=32\times 32$ with the periodic boundary condition.

Regarding the impurity solver, we used hybridization expansion algorithm of the continuous-time quantum Monte Carlo (CTQMC) method~\cite{Werner06a, Gull11} in the region $U/W \lesssim 1.5$.
In the strong-coupling regime $U/W\gtrsim 1.0$, we used the Hubbard-I approximation to obtain a stable solution, avoiding statistical errors (see Appendix~\ref{app:Hubbard-I} for some remarks).
In solving the matrix equation in Eq.~(\ref{eq:chiq-DMFT}), we introduced a cutoff for the fermionic frequencies.
See Appendix~\ref{app:freq_cutoff} for the technical details.

\begin{figure}[tb]
	\begin{center}
	\includegraphics[width=0.9\linewidth]{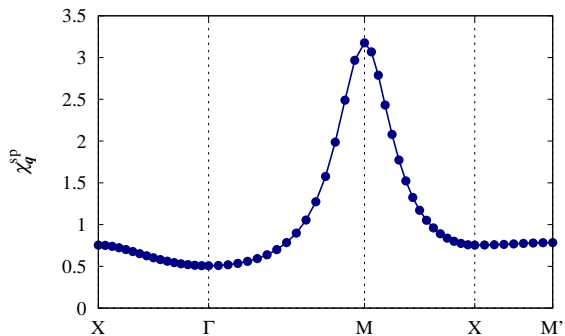}
	\end{center}
	\caption{Momentum dependence of the static spin susceptibility $\chiq^{\rm sp}$ for the square-lattice Hubbard model with $U/W=1$ and $T=0.5$. The CTQMC solver was used. $\Gamma$: (0,0), X: (0,$\pi$), M: ($\pi$,$\pi$) $\equiv \bm{Q}$, $M'$: ($\pi/2$,$\pi/2$).}
	\label{fig:chiq_path}
\end{figure}

\begin{figure}[tb]
	\begin{center}
	\includegraphics[width=\linewidth]{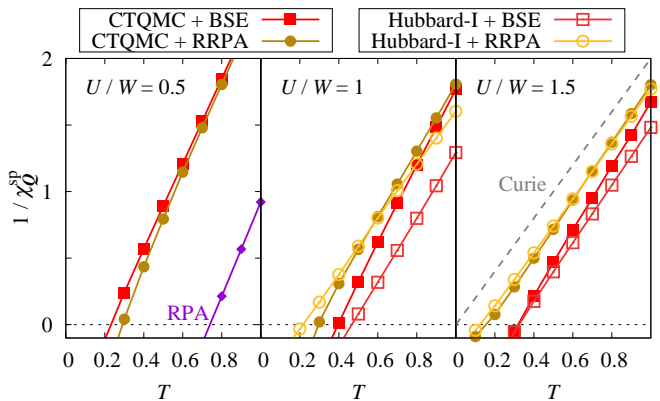}
	\end{center}
	\caption{Temperature dependence of the inverse susceptibility, $1/\chiQ^{\rm sp}$, for the AFM fluctuation in the Hubbard model on a half-filled square lattice. The closed (open) symbols are CTQMC (Hubbard-I) results. The squares are results computed with the original BS equation in Eq.~(\ref{eq:chiq-DMFT}), and circles are results computed with RRPA in Eq.~(\ref{eq:chiq-rrpa}).}
	\label{fig:T-chiq}
\end{figure}

\begin{figure}[tb]
	\begin{center}
	\includegraphics[width=\linewidth]{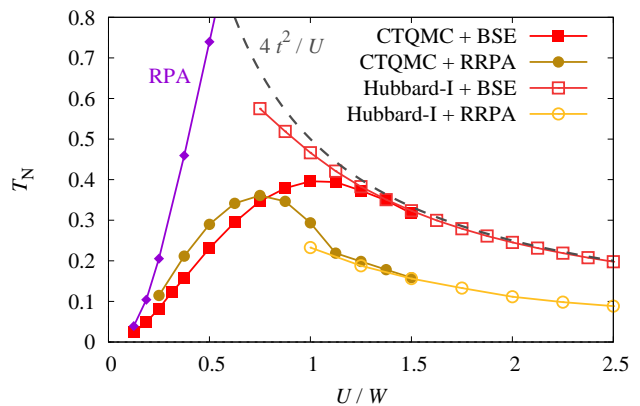}
	\end{center}
	\caption{Phase diagram of the Hubbard model on a square lattice. See the caption of Fig.~\ref{fig:T-chiq} for explanations of symbols.}
	\label{fig:phase_rrpa}
\end{figure}

Figure~\ref{fig:chiq_path} shows the momentum dependence of the static spin susceptibility $\chiq^\textrm{sp}$.
Here, $\chiq^{\rm sp}$ is defined as the correlation between the operator
$O^\textrm{sp}=(O_{\uparrow\uparrow} - O_{\downarrow\downarrow})/\sqrt{2}$.
From this figure, we confirm that fluctuations at $\bm{q}=(\pi,\pi) \equiv \bm{Q}$ is the strongest.
Then, we fix $\bm{q}$ at $\bm{Q}$ and look into the $T$ dependence of $\chiQ^{\rm sp}$.
Figure~\ref{fig:T-chiq} shows the inverse of $\chiQ^{\rm sp}$ as a function of $T$ for $U/W=0.5$, 1, and 1.5 (red squares).
In the DMFT, $\chiQ^{\rm sp}$ basically follows the Curie-Weiss law, $\chiQ^{\rm sp} \simeq C/(T-T_\textrm{N})$, due to the mean-field nature of the intersite correlations\footnote{In other words, the two-particle sum-rule $N^{-1}\sum_{\bm{q}} \chi_{\bm{q}}=\chi_\textrm{loc}$ is not fulfilled in the DMFT. To recover it, spatial correlations need to be included. See Ref.~\onlinecite{Rohringer18} for a review of such extended theories.}.
The Curie constant $C$ agrees with that estimated from the localized spin, $C=1/2$, in the strong-coupling limit, and is reduced in the weak-coupling regime.
The transition temperature determined from the divergence of $\chiQ^{\rm sp}$ is plotted as a function of $U$ in Fig.~\ref{fig:phase_rrpa}.
Convergence to the Heisenberg limit $T_\textrm{N} \sim 4t^2/U$ is confirmed.

Two reference results are shown in Figs.~\ref{fig:T-chiq} and \ref{fig:phase_rrpa} for comparison.
One is the RPA computed by
\begin{align}
(\chiq^\textrm{RPA})^{-1} = (\chi_{0,\bm{q}}^{U=0})^{-1} - U.
\label{eq:chi-rpa}
\end{align}
Here, $\chi_{0,\bm{q}}^{U=0}$ denotes the susceptibility for $U=0$ (no self-energy correction).
The RPA yields $\TN$ that increases monotonically with increasing $U/W$ , and hence $\TN$ is considerably overestimated in the whole parameter region, as expected.
Improvement to the RPA susceptibility is made by including correlations from DMFT calculations.
We consider the following quantity, which we refer to as renormalized RPA (RRPA):
\begin{align}
(\chiq^\textrm{RRPA})^{-1} = \chi_{0,\bm{q}}^{-1} - \chi_\textrm{0,loc}^{-1} + \chiloc^{-1}.
\label{eq:chiq-rrpa}
\end{align}
This expression corresponds to replacing all $X$ with $\chi$ in Eq.~(\ref{eq:chiq-DMFT})
(namely, summation over the fermionic Matsubara frequencies are taken independently on each $X$).
The RRPA formula takes two corrections into account compared with that in RPA:
Firstly, $G$ in $\chi_0$ includes the self-energy correction, and secondly, the bare interaction $U$ is replaced with the effective one,
$U_\textrm{eff}=\chi_{0,\rm loc}^{-1} - \chiloc^{-1}$.
Figure~\ref{fig:phase_rrpa} demonstrates that the RRPA yields a reasonable value between the exact and RPA for $U \lesssim W$.
However, it underestimates $\TN$ for $U \gtrsim W$ and does not gives the correct asymptotics for large $U$.
Consequently, the RRPA results in a wrong phase diagram in which some phase is missing (orb$^z$) in multiorbital cases (see Sec.~\ref{sec:results-multiorb} for an explicit result).
In the rest of this paper, we propose an alternative approach which is valid in the strong-coupling regime.
The formula is as simple as the RPA formula and is executable with a low numerical cost comparable to the RRPA.

\section{Strong-coupling formula}
\label{sec:scl}
In this section, we derive a simple susceptibility formula which gives the correct asymptotics in the large-$U$ limit.
To avoid complexity and illustrate our main idea clearly, we restrict ourselves to a single-band model in this section. 
Extension to multiorbital systems will be presented in Sec.~\ref{sec:multiorbital}.

\subsection{Decoupling}
\begin{figure}[tb]
	\begin{center}
	\includegraphics[width=0.48\linewidth]{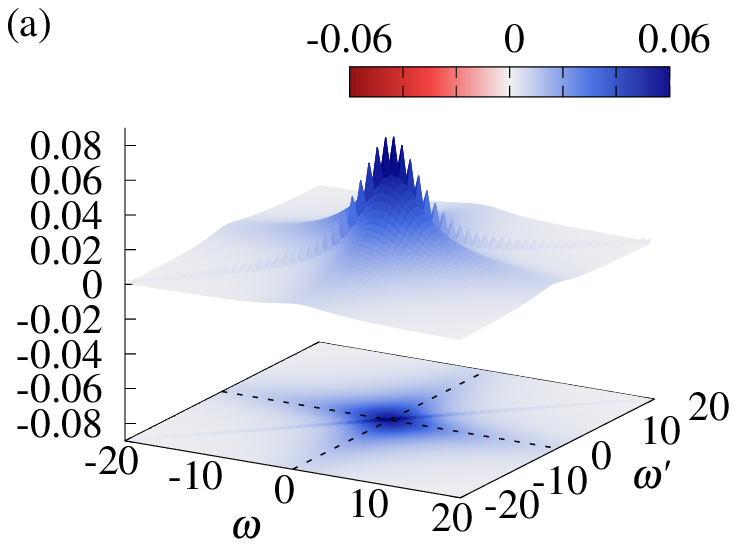}
	\includegraphics[width=0.48\linewidth]{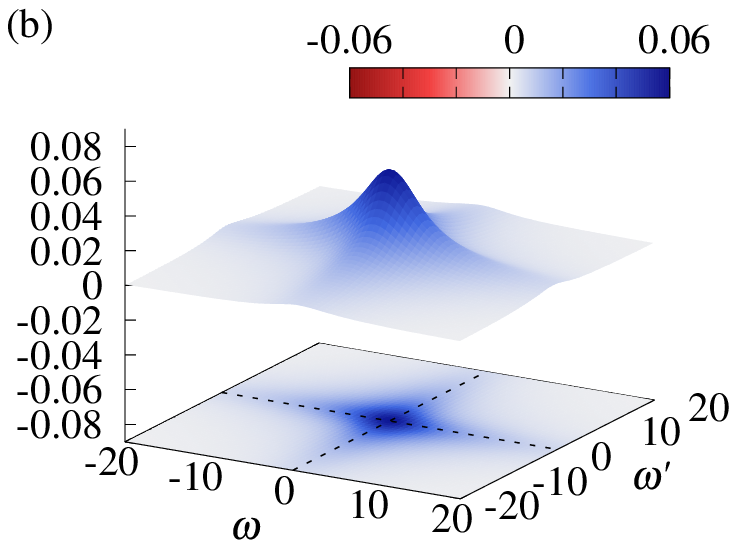}
	\includegraphics[width=0.48\linewidth]{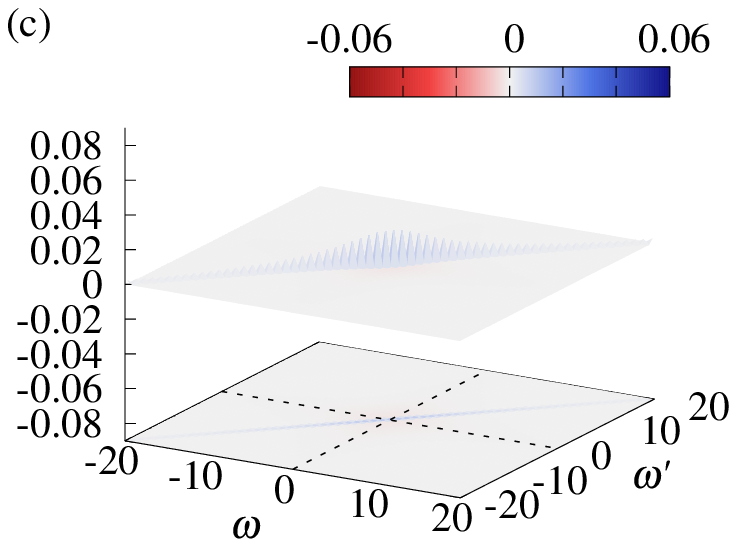}
	\includegraphics[width=0.48\linewidth]{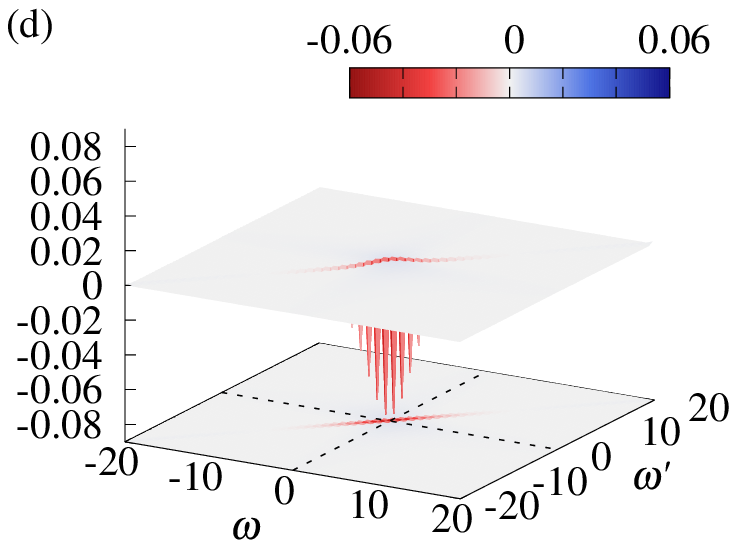}
	\end{center}
	\caption{(a) The two-particle Green function in the spin channel, $\textrm{Re}\Xloc^\textrm{sp}(\iw,\iw')$, in the Hubbard model with $U/W=2$ and $T=0.1$ computed by the Hubbard-I approximation. (b) The result after the decoupling approximation in Eq.~(\ref{eq:X-SVD}). (c) The difference between (a) and (b). (d) The charge channel $\textrm{Re}\Xloc^\textrm{ch}(\iw,\iw')$.}
	\label{fig:Xloc}
\end{figure}

In Eq~(\ref{eq:chiq-DMFT}), the frequency dependence in $\Xloc(\iw,\iw')$ is particularly important in the strong coupling regime.
The RRPA formula filters out 
the frequency dependence over $(\iw, \iw')$ space
and hence fails in reproducing the correct results for $U \gtrsim W$.
Let us begin with observing $\Xloc(\iw,\iw')$ to understand its structure in the $\omega$-$\omega'$ plane.
Figure~\ref{fig:Xloc}(a) shows numerical results for $\Xloc(\iw,\iw')$ in the spin channel (we represent it as $\Xloc^\textrm{sp}$).
It consists of sharp contribution at the diagonal, $\omega=\omega'$, and a broad peak around the origin, $\omega=\omega'=0$.

Our finding is that the broad contribution describes fluctuations of local degrees of freedom, which are relevant for $\chiq$, and the diagonal components are irrelevant.
Furthermore, the broad peak in the $\omega$-$\omega'$ plane can be decoupled as
\begin{align}
\Xloc(\iw, \iw') \simeq \Phi(\iw) \Phi(\iw').
\label{eq:X-decouple}
\end{align}
Mathematically, this approximation is carried out by considering $\Xloc(\iw, \iw')$ as a matrix with respect to $\omega$ and $\omega'$, and performing the singular value decomposition (SVD) to retain only the leading term:
\begin{align}
\Xloc(\iw, \iw') &= \sum_{i\geq0} s_i u_i(\iw) v_i^*(\iw')
\simeq s_0 u_0(\iw) v_0^*(\iw').
\label{eq:X-SVD}
\end{align}
Here, $\{ u_i(\iw) \}$ and $\{ v_i(\iw) \}$ each is an orthogonal set and normalized by
$\sum_{\omega} |u_i(\iw)|^2 = \sum_{\omega} |v_i(\iw)|^2 = 1$.
Provided that the phase factor of $u_0(\iw)$ and $v_0^*(\iw)$ are properly chosen, those functions satisfy $u_0(\iw)=v_0^*(\iw)$ (for details, see Appendix~\ref{app:u0-v0}).
Thus, $\Phi(\iw)$ can be obtained by
\begin{align}
\Phi(\iw)=\sqrt{s_0}u_0(\iw)=\sqrt{s_0}v_0^*(\iw).
\end{align}

\begin{figure}[tb]
	\begin{center}
	\includegraphics[width=\linewidth]{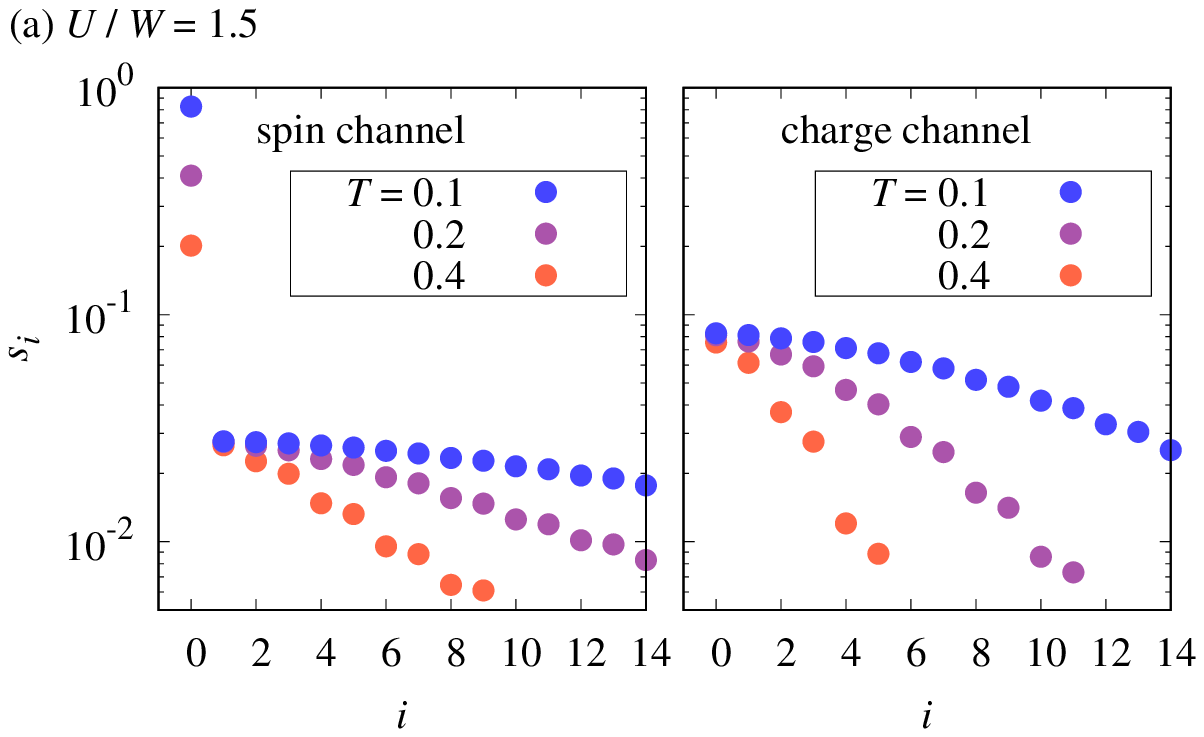}
	\includegraphics[width=\linewidth]{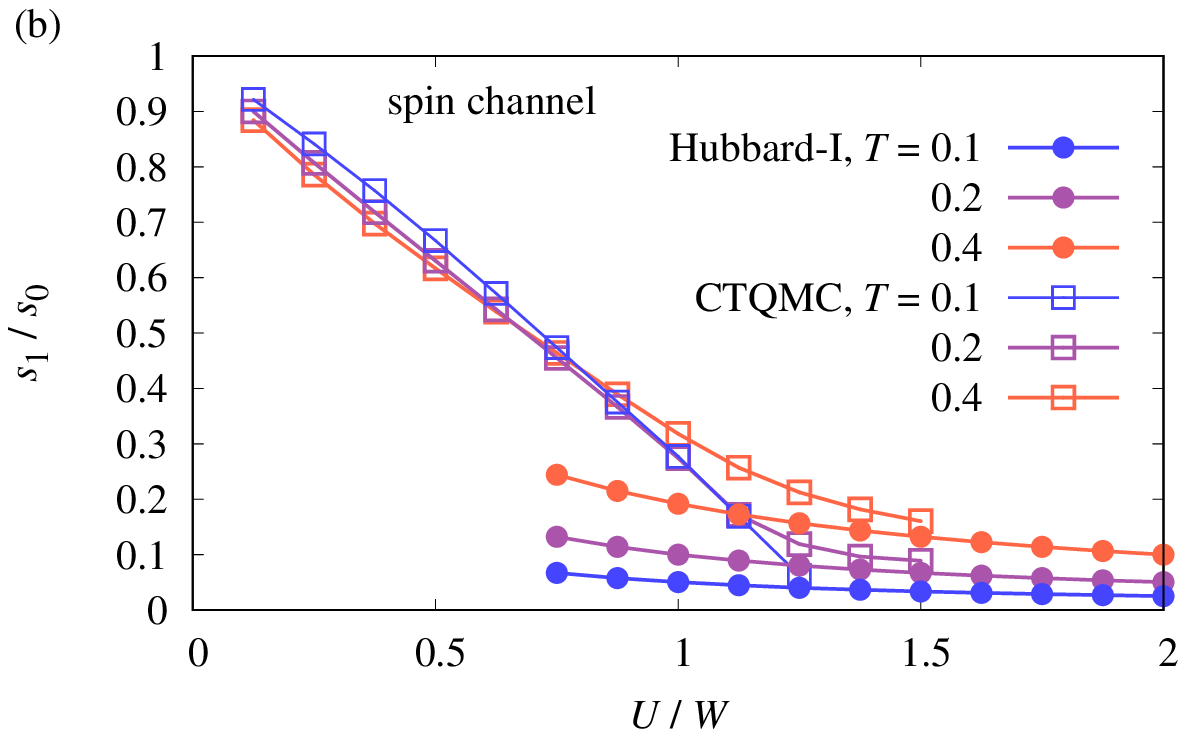}
	\end{center}
	\caption{(a) The singular values $s_i$ of $\Xloc(\iw,\iw')$ for the spin channel (left) and for the charge chance (right). The half-filled Hubbard model with $U/W=1.5$.
	(b) The ratio $s_1/s_0$ for the spin channel as a function of $U$.}
	\label{fig:SV}
\end{figure}

The left panel of Fig.~\ref{fig:SV}(a) shows the singular values $s_i$ of $\Xloc^\textrm{sp}(\iw,\iw')$ in the Hubbard model.
It turns out that the first component $s_0$ is larger than the second component $s_1$ by 1--2 orders of magnitude.
We found that $s_0$ increases in proportion to $1/T$.
The ratio $s_1/s_0$ plotted as a function of $U$ in Fig.~\ref{fig:SV}(b) demonstrates a rapid reduction of the higher-order contribution as $U$ increases and $T$ decreases.
We conclude from these results that the decoupling in Eq.~(\ref{eq:X-SVD}) is valid in the strong-coupling and low-temperature regime.

After the decoupling, $\Xloc^\textrm{sp}(\iw,\iw')$ in Fig.~\ref{fig:Xloc}(a) is approximated into Fig.~\ref{fig:Xloc}(b), and Fig.~\ref{fig:Xloc}(c) shows the neglected part.
It is clear from these figures that the broad peak is extracted by $s_0$,
and the diagonal components are neglected, demonstrating that our central idea is supported in terms SVD.

For comparison, the charge channel, $\Xloc^\textrm{ch}(\iw,\iw')$, is plotted in Fig.~\ref{fig:Xloc}(d).
Here, the charge operator is defined by $O^\textrm{ch}=(O_{\uparrow\uparrow} + O_{\downarrow\downarrow})/\sqrt{2}$.
In contrast to the spin channel, the dominant structure appears to be the diagonal component, and a broad peak around $\omega=\omega'=0$ is missing.
In this case, the singular values $s_i$ vary continuously as shown in the right panel of Fig.~\ref{fig:SV}(a).
The decoupling approximation, therefore, does not work for the charge channel.

The contrasting behaviors of $s_i$ in the spin and charge channels indicate that the singular peak at $s_0$ is indicative of residual degrees of freedom in the ground state.
As the local spin fluctuations get stronger according to the Curie law, the singularity in $s_0$ becomes more conspicuous.
Since long-range order takes place in such situations, the decoupling approximation is expected to be suitable for descriptions of phase transitions.

Let us finally derive an explicit relation between $s_0$ and $\chiloc$.
For this purpose, we use Eq.~(\ref{eq:chi_and_X}) as a ``sum-rule" for $\Xloc(\iw,\iw')$.
Substituting Eq.~(\ref{eq:X-decouple}) into Eq.~(\ref{eq:chi_and_X}), we obtain
\begin{align}
T \chiloc \simeq \overline{\Phi}^2 = s_0 \overline{u}_0^2,
\label{eq:s0-sumrule}
\end{align}
where the overline signifies an average defined by
$\overline{\Phi} \equiv T\sum_{\omega} \Phi(\iw) e^{\iw 0^+}$.
At low-$T$, $\overline{u}_0^2$ is proportional to $T$ (see Sec.~\ref{sec:u0}), and we obtain $s_0 \propto \chiloc$.

\subsection{The SCL formula}

Now, we apply the decoupling in Eq.~(\ref{eq:X-decouple}) to the BS equation in Eq.~(\ref{eq:chiq-DMFT}).
To this end, we first rewrite Eq.~(\ref{eq:chiq-DMFT}) in the form which is suitable in the strong-coupling regime:
\begin{align}
\bXq^{-1} = \bXloc^{-1} - \bPq,
\label{eq:chiq-DMFT-2}
\end{align}
where
\begin{align}
\bPq \equiv \bXlocZero^{-1} - \bXqZero^{-1}.
\label{eq:P_q}
\end{align}
Equation~(\ref{eq:chiq-DMFT-2}) differs from the weak-coupling formula like the RRPA in the sense that 
the starting point is the local susceptibility and the non-local correction is taken into account by $\bPq$.
Expanding the right-hand side of Eq.~(\ref{eq:chiq-DMFT-2}) requires evaluation of the product
$\bXloc \bPq \bXloc$, which can be evaluated under the decoupling approximation as
\begin{align}
\sum_{\omega''} \Xloc(\iw, \iw'') \Pq(\iw'') \Xloc(\iw'', \iw')
\simeq
\lambda_{\bm{q}} \Xloc(\iw,\iw'),
\end{align}
where we defined $\lambda_{\bm{q}}$ by
\begin{align}
\lambda_{\bm{q}} \equiv \sum_{\omega''} \Pq(\iw'') \Phi(\iw'')^2.
\label{eq:lambda_q-1}
\end{align}
Using this result, Eq.~(\ref{eq:chiq-DMFT-2}) can be expressed as
$\bXq \simeq (1-\lambda_{\bm{q}})^{-1} \bXloc$.
The summations over $\omega$ and $\omega'$ can be now taken explicitly to yield
\begin{align}
\chiq^\textrm{SCL} = (1-\lambda_{\bm{q}})^{-1} \chiloc.
\label{eq:chiq-SCL-lambda}
\end{align}
Here, the superscript SCL stands for strong-coupling limit.
Equation~(\ref{eq:chiq-SCL-lambda}) is our susceptibility formula derived in the strong-coupling limit.

It is convenient to express Eq.~(\ref{eq:chiq-SCL-lambda}) in 
an RPA-like form.
Defining a quantity $\Iq$ by
\begin{align}
\Iq \equiv \lambda_{\bm{q}}/\chiloc,
\label{eq:I_q-original}
\end{align}
Eq.~(\ref{eq:chiq-SCL-lambda}) is rewritten as
\begin{align}
\chiq^\textrm{SCL} = (\chiloc^{-1} - \Iq)^{-1}.
\label{eq:chiq-SCL}
\end{align}
One can see that $\Iq$ represents effective intersite interactions, which give rise to the correction to the local susceptibility, $\chiloc$.
In the next section (Sec.~\ref{sec:analytical}), we will show that, indeed, $\Iq$ correctly reproduces the kinetic exchange and RKKY interactions in the Hubbard and periodic Anderson models, respectively. 

Finally, it is important to confirm, in Eq.~(\ref{eq:I_q-original}), that $\chiloc$ is canceled with $\lambda_{\bm{q}}$ and $\Iq$ is independent of $\chiloc$
(otherwise, $\Iq$ does not make physical sense).
To see this, we use the relation in Eq.~(\ref{eq:s0-sumrule}), which holds in the strong-coupling limit, and express $\Iq$ as
\footnote{Actually, this expression overestimates the interactions because Eq.~(\ref{eq:s0-sumrule}) holds only in the strong-coupling limit and generally we have $T\chiloc > s_0 \overline{u}_0^2$. Nevertheless, we will use this expression in the SCL3 scheme (Sec.~\ref{sec:SCL3}) because no estimation of $s_0$ is necessary.}
\begin{align}
&\Iq \simeq T \sum_{\omega} \Pq(\iw) \phi(\iw)^2,
\label{eq:I_q}
\end{align}
where we introduced a normalized function $\phi(\iw)=\Phi(\iw)/\overline{\Phi}$.
It is now clear that, since $\phi(\iw)$ is normalized, $\Iq$ does not depend on the magnitude of local fluctuations.

\subsection{Numerical verification of the SCL formula}
\label{sec:numerical}

\begin{figure}[tb]
	\begin{center}
	\includegraphics[width=\linewidth]{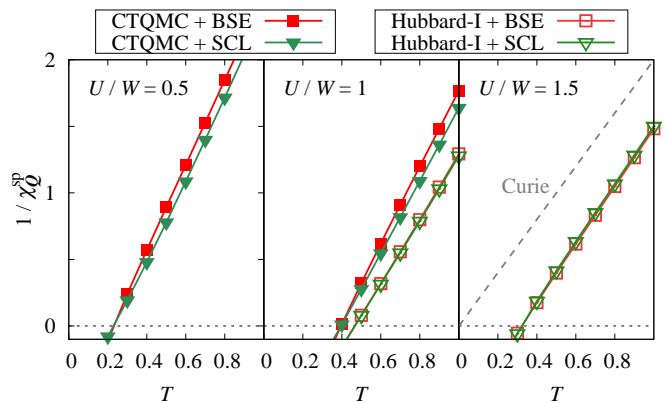}
	\end{center}
	\caption{Comparison of $1/\chiQ^{\rm sp}$ computed in two ways. The green triangles are results computed by the SCL formula, Eq.~(\ref{eq:chiq-SCL}), and using the SVD (termed as SCL1 scheme later). The red squares are those computed by the original BS equation (the same data as in Fig.~\ref{fig:T-chiq}).}
	\label{fig:T-chiq-SCL}
\end{figure}

Let us apply the SCL formula, Eq.~(\ref{eq:chiq-SCL}), to the Hubbard model to check the validity of the approximation.
Figure~\ref{fig:T-chiq-SCL} shows a comparison between $\chiq^\textrm{SCL}$ and $\chiq$ computed with the original BS equation.
For the strong-coupling parameter, $U/W=1.5$, the two results agree quite well as expected.
For weaker coupling, on the other hand, 
$\chiq^\textrm{SCL}$ is larger than $\chiq$, namely, overestimate fluctuations at high $T$.
However, the deviation gets smaller as $T$ decreases,
and the two results coincide at the transition temperature.
This result indicates that the decoupling approximation does not cause a loss in quality of estimation of the transition.

\begin{figure}[tb]
	\begin{center}
	\includegraphics[width=\linewidth]{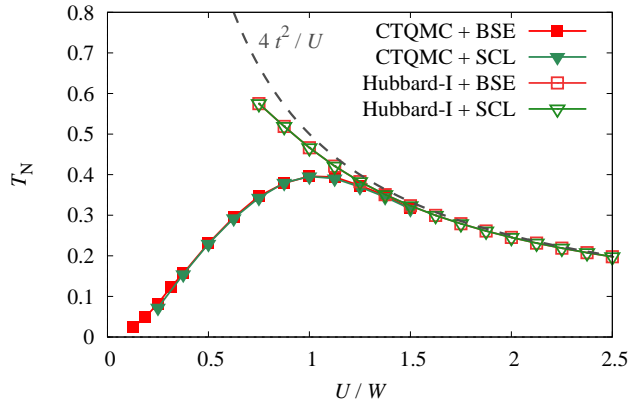}
	\end{center}
	\caption{Comparison of $T_\textrm{N}$ computed by the SCL formula (termed as SCL1 scheme later) and that computed by the original BS equation.}
	\label{fig:phase_scl}
\end{figure}

The N\'eel temperature $T_\textrm{N}$ is shown in Fig.~\ref{fig:phase_scl} as a function of $U/t$.
Surprisingly, our formula, which was derived using the strong-coupling features of $\Xloc(\iw, \iw')$, shows complete agreement with the correct $T_\textrm{N}$ even in the weak-coupling regime down to $U/W = 2/8$.
From this result, we conclude that our SCL formula, Eq.~(\ref{eq:chiq-SCL}), provides the correct estimation of the transition temperature,
though the fluctuations are overestimated for $T>T_\textrm{N}$.
Multiorbital models will be examined in Sec.~\ref{sec:results-multiorb}.

\begin{figure}[tb]
	\begin{center}
	\includegraphics[width=0.9\linewidth]{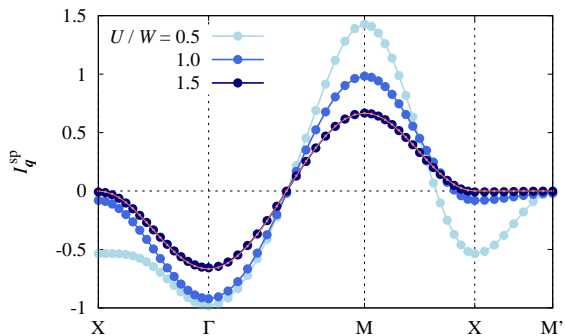}
	\end{center}
	\caption{Momentum dependence of the the effective exchange interactions $\Iq^\textrm{sp}$ computed by the SCL formula for $T=0.5$. The lines with a dot show the analytic expression $\Iq^\textrm{atom}$ for $U/W=1.5$.}
	\label{fig:Iq_path}
\end{figure}

\begin{figure}[tb]
	\begin{center}
	\includegraphics[width=0.85\linewidth]{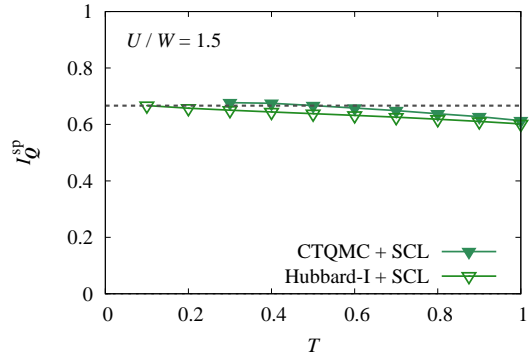}
	\end{center}
	\caption{$T$-dependence of $\IQ^\textrm{sp}$ for $U/W=1.5$. The dashed line indicates the atomic-limit value $\IQ^\textrm{atom}=2/3$.}
	\label{fig:T-Iq}
\end{figure}

We close this section by discussing the effective nonlocal interactions, $\Iq$, which can be derived using the SCL formula by virtue of the simplified treatment of the BS equation.
Figure~\ref{fig:Iq_path} shows $\Iq^\textrm{sp}$ in the spin channel computed by Eq.~(\ref{eq:I_q-original}).
The positive (negative) values indicate enhancement (suppression) of the fluctuations, and the peak at M-point ($\bm{q}=\bm{Q}$) means antiferromagnetic interactions.
The result for $U/W=1.5$ perfectly agrees with the analytic formula in the atomic limit, $\Iq^\textrm{atom} = -(4t^2/U) (\cos k_x + \cos k_y)$, demonstrating the validity of the SCL formula.
This will be proved analytically in the next section.
For smaller $U$, the result shows deviation from the simple cosine function (especially at X-point), which indicates that the second-neighbor or even longer-range interactions are effectively active.
Figure~\ref{fig:T-Iq} shows the $T$-dependence of $\IQ^\textrm{sp}$. It turns out that the value only weakly depends on $T$.
Thus, it has been confirmed that $\Iq$ can be regarded as a physical effective interaction between the localized spins.

\section{Atomic limit of the SCL formula}
\label{sec:analytical}

Our SCL formula in Eq.~(\ref{eq:chiq-SCL}) is expected to capture correct physics in the strong-coupling regime.
In this section,
we verify analytically that the formula yields the well-known effective interactions emerging in the strong-coupling limit of two fundamental lattice models:
the kinetic exchange interaction in the Hubbard model and the RKKY interaction in the periodic Anderson model.

\subsection{The function $u_0(\iw)$ in the atomic limit}
\label{sec:u0}

The expression for $\Iq$ in Eq.~(\ref{eq:I_q}) includes two kinds of contribution: 
$\Pq(\iw)$ takes account of the energy-band structure, and $\phi(\iw)$ or $u_0(\iw)$ takes local correlations into account.
We first clarify the features of the local quantities in this subsection, and then derive the effective interactions in two lattice models in the succeeding subsections.

\begin{figure}[tb]
	\begin{center}
	\includegraphics[width=\linewidth]{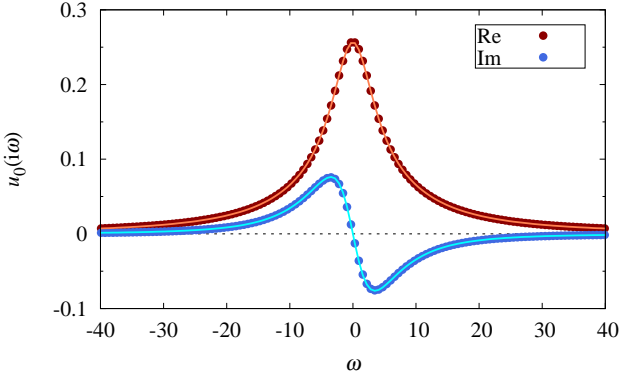}
	\end{center}
	\caption{The function $u_0(\iw)$ for a single-orbital Hubbard atom with $U=16$, $\mu=4$, and $T=0.1$. The lines show the fitting results using Eq.~(\ref{eq:u0-atom}).}
	\label{fig:u0}
\end{figure}

Figure~\ref{fig:u0} shows the numerical result for $u_0(\iw)$ in a single-orbital atom with the Hubbard interaction.
Here, the phase of $u_0(\iw)$ is fixed so that $\sum_{\omega} u_0(\iw)$ is positive real.
The parameters $U$ and $\mu$ are chosen so that the ground state is the singly-occupied states without the particle-hole symmetry.
We found 
numerically
that $u_0(\iw)$ at low temperatures can be perfectly fitted
by the two-pole function of the form
\begin{align}
u_0^\textrm{atom}(\iw) = \frac{A}{2} \left( \frac{1}{\iw+\mu} - \frac{1}{\iw+\mu-U} \right),
\label{eq:u0-atom}
\end{align}
where $A$ is a real normalization factor determined by the condition $\sum_{\omega} |u_0^\textrm{atom}(\iw)|^2=1$.
The form of Eq.~(\ref{eq:u0-atom}) can be interpreted in terms of single-particle excitations:
The first contribution is due to the excitation to empty state, and the second due to the excitation to doubly occupied state.
In the case with the particle-hole symmetry, namely $\mu=U/2$, the function $u_0^\textrm{atom}(\iw)$ is reduced to a Lorentzian of the form
$u_0^\textrm{atom}(\omega) = A (U/2)[\omega^2 + (U/2)^2]$.

For the sake of later use, we derive explicit expressions of $A$ and $s_0$.
For simplicity, we consider the low-temperature region, i.e., $T \ll \mu$ and $T \ll U-\mu$, in which the average $\overline{u}_0$, for example, is simplified into $\overline{u}_0 \simeq A/2$.
From the normalization condition of $u_0^\textrm{atom}(\iw)$, $A$ is evaluated as
\begin{align}
A^2 \simeq 8UT r(1-r),
\label{eq:A2}
\end{align}
where $r \equiv \mu/U$.
Furthermore, using the Curie law $\chiloc=1/2T$ ($1/2Tn_\textrm{orb}$ in the case with orbital degeneracy) in Eq.~(\ref{eq:s0-sumrule}), we obtain
\begin{align}
s_0 \simeq \frac{1}{4UTr(1-r)}.
\end{align}
This result is consistent with Fig.~\ref{fig:SV}(a), which exhibits an increase of $s_0$ with decreasing $T$.

Finally, the function $\phi(\iw)$ defined in connection with Eq.~(\ref{eq:I_q}) is evaluated by $\phi(\iw)=u_0(\iw)/\overline{u}_0$, which becomes
\begin{align}
\phi^\textrm{atom}(\iw) = \frac{1}{\iw+\mu} - \frac{1}{\iw+\mu-U}.
\label{eq:phi-atom}
\end{align}

\subsection{Kinetic exchange interaction in the Hubbard model}
The Hubbard model is reduced to the Heisenberg model in the strong-coupling limit. The exchange coupling constant is proportional to $t^2/U$.
Applying Eq.~(\ref{eq:chiq-SCL}) to the Hubbard model, we demonstrate below that $\chiq^\textrm{SCL}$ corresponds to the mean-field susceptibility of the effective Heisenberg model.

\begin{figure}[tb]
	\begin{center}
	\includegraphics[width=0.95\linewidth]{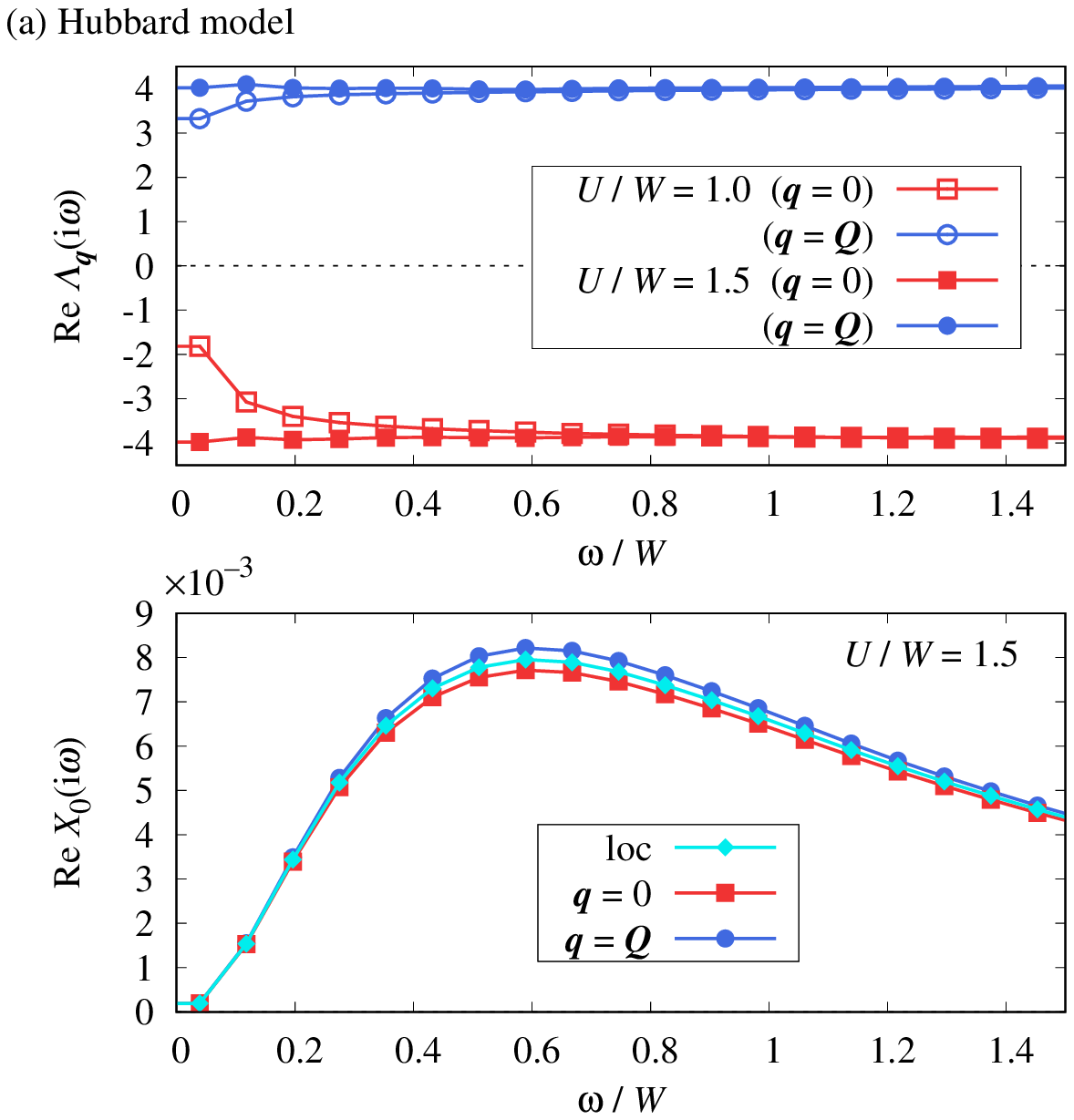}
	\includegraphics[width=0.95\linewidth]{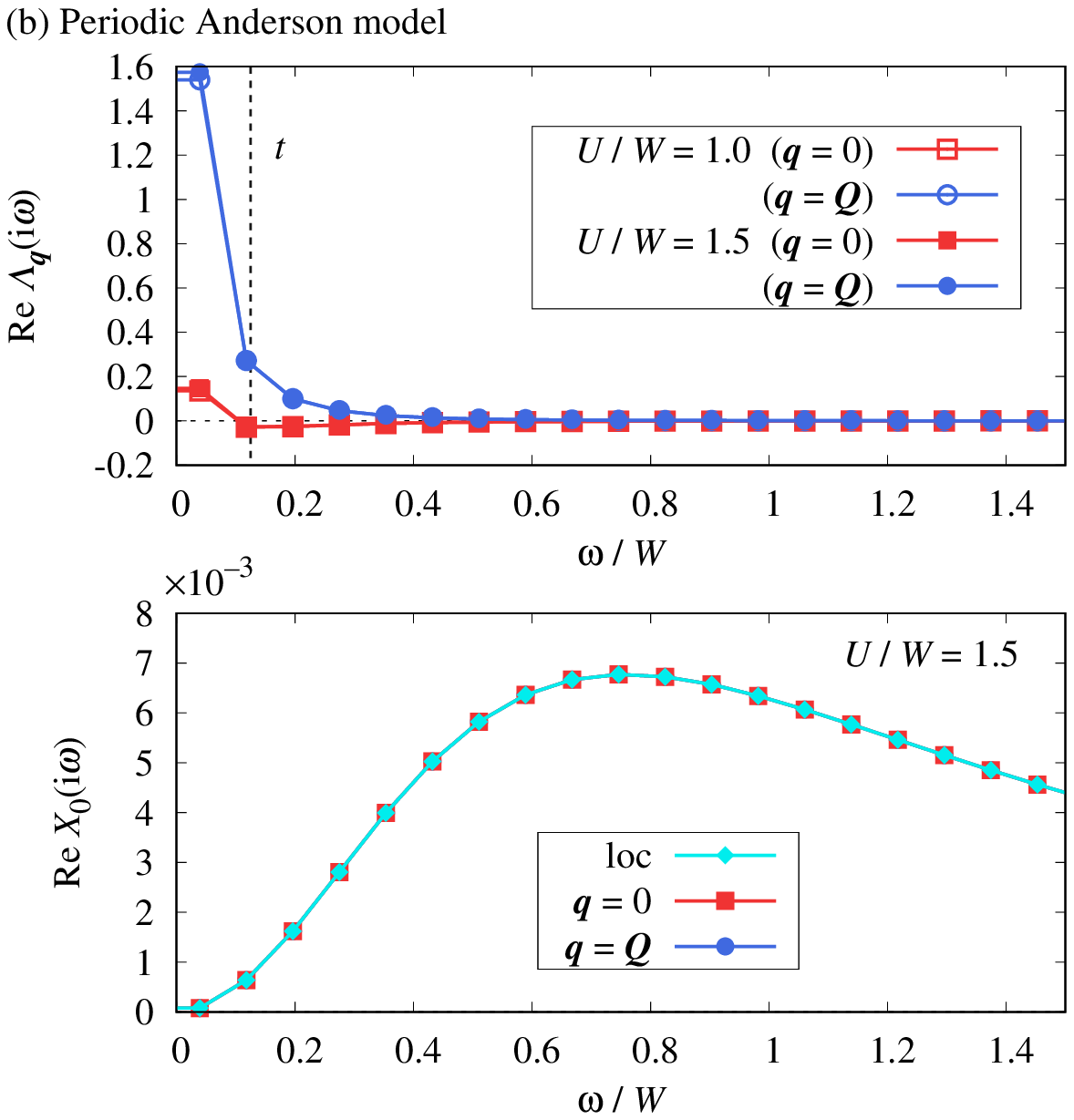}
	\end{center}
	\caption{$\Pq(\iw)$ (upper panel) and $X_{0,\bm{q}}(\iw)$ and $X_{0,\textrm{loc}}(\iw)$ (lower panel) as a function of $\omega$. (a) The Hubbard model with $n=1$, $T=0.1$, and (b) the periodic Anderson model with $V^2=1$, $\epsilon_f=-U/2$, $T=0.1$. The red squares are $\bm{q}=0$ and the blue circles are $\bm{q}=\bm{Q}$. The impurity solver used is the CTQMC method in (a) and the Hubbard-I approximation in (b).
	}
	\label{fig:P_q}
\end{figure}

A key observation is that the quantity $\Pq(\iw)$ defined in Eq.~(\ref{eq:P_q}) becomes independent of $\omega$ in the strong-coupling limit.
Given that the energy dispersion $\epsilon_{\bm{k}}$ takes the form
$\epsilon_{\bm{k}}= 2t\gamma_{\bm{k}}$ with $\gamma_{\bm{k}} \equiv \cos k_x + \cos k_y$,
the quantity $\Pq(\iw)$ becomes
\begin{align}
\Pq(\iw) \simeq -2t^2 \gamma_{\bm{q}}.
\label{eq:Pq-Hubbard}
\end{align}
The detailed derivation is presented in Appendix~\ref{app:Pq-Hubbard}.
The numerical result in Fig.~\ref{fig:P_q}(a) shows that $\Pq(\iw)$ is well represented by Eq.~(\ref{eq:Pq-Hubbard}) for $U/W=1.5$.
We note that the quantities $X_{0,\bm{q}}(\iw)$ and $X_{\textrm{0,loc}}(\iw)$ [see Eq.~(\ref{eq:P_q})] do depend on $\omega$ 
[Fig.~\ref{fig:P_q}(a)], however, the $\omega$-dependence cancels out in  $\Pq(\iw)$.

Substitution of Eqs.~(\ref{eq:Pq-Hubbard}) and (\ref{eq:phi-atom}) into Eq.~(\ref{eq:I_q}) yields the coupling constant $\Iq$ of the form
\begin{align}
\Iq \simeq -\frac{4t^2 \gamma_{\bm{q}}}{U},
\end{align}
where we used $A^2=2UT$, assuming
the half-filling, $\mu = U/2$.
This expression corresponds to the kinetic exchange interaction between neighboring sites.
The transition temperature is determined by the condition $\chiloc \Iq=1$.
Using $\Iq=8t^2/U$ and the Curie law $\chiloc=1/2T$, we obtain the N\'eel temperature $T_\textrm{N}=4t^2/U$, which is consistent with the Heisenberg limit of the Hubbard model.

\subsection{RKKY interaction in the periodic Anderson model}
We next consider the periodic Anderson model and derive the RKKY interaction between localized electrons. 
The Hamiltonian consists of band electrons written with the $c$ operator and localized (correlated) electrons written with $f$ operators:
\begin{align}
\mathcal{H} &= \sum_{\bm{k}\sigma} \epsilon_{\bm{k}} c_{\bm{k}\sigma}^{\dag} c_{\bm{k}\sigma}
+ \epsilon_f \sum_{\bm{k}\sigma} f_{\bm{k}\sigma}^{\dag} f_{\bm{k}\sigma}
\nonumber \\
&+ V \sum_{\bm{k}} \left( f_{\bm{k}\sigma}^{\dag} c_{\bm{k}\sigma} + c_{\bm{k}\sigma}^{\dag} f_{\bm{k}\sigma} \right)
+U \sum_{i} n_{fi\uparrow} n_{fi\downarrow},
\label{eq:H-PAM}
\end{align}
where $n_{fi\sigma}=f_{i\sigma}^{\dag} f_{i\sigma}$.
The ordinary derivation of the RKKY interaction is given as the fourth-order perturbation with respect to $V$ (see for example Ref.~\onlinecite{Shiba99}).
We will derive a similar result based on the DMFT, starting from the SCL formula.

We consider the susceptibility $\chi(\bm{q})$ of $f$ electrons.
Although there are two bands ($c$ and $f$) in this model, the BS equation is closed only with $f$ quantities because no two-body interaction is supposed between $c$ electrons~\cite{Jarrell95}.
Hence, we simply regard all equations in Sec.~\ref{sec:scl} as those for $f$-components.

In the atomic limit of $\Iq$ in Eq.~(\ref{eq:I_q}),
$\phi(\iw)$ is common between the Hubbard model and the periodic Anderson model.
Hence, the difference between the two models arises from $\Pq(\iw)$.
In the strong-coupling limit, the analytical expression of $\Pq(\iw)$ reads
\begin{align}
\Pq (\iw) = V^4 [ X_{\textrm{c},\bm{q}} (\iw) - \bar{X}_\textrm{c} (\iw) ],
\label{eq:Pq-PAM}
\end{align}
where
\begin{align}
X_{\textrm{c},\bm{q}} (\iw) &= -\langle g_{\textrm{c},\bm{k}}(\iw) g_{\textrm{c},\bm{k}+\bm{q}}(\iw) \rangle_{\bm{k}}, \\
\bar{X}_{\textrm{c}} (\iw) &= -\bar{g}_{\textrm{c}}(\iw) \bar{g}_{\textrm{c}}(\iw).
\end{align}
The detailed derivation is presented in Appendix~\ref{app:Pq-PAM}.
Figure~\ref{fig:P_q}(b) shows $\Pq(\iw)$ in the periodic Anderson model, and demonstrates a striking difference with that in the Hubbard model.
We note that no qualitative difference is seen in $X_{\textrm{0}, \bm{q}}(\iw)$, confirming that the quantity $\Pq(\iw)$ is the relevant quantity which characterizes each model.
In Fig.~\ref{fig:P_q}(b), we remark that $\Pq(\iw)$ decays
rapidly above $\omega \sim t$ [because $t$ is the only energy scale that enters in $g_{c,\bm{k}}(\iw)$].
On the other hand, 
$\phi(\iw)$
shows the power-law decay with the length $\omega \sim U$ [see Eq.~(\ref{eq:phi-atom}) and Fig.~\ref{fig:u0}].
Therefore, we can replace the $\omega$-dependence in $\Pq(\iw)$ with $\delta_{\omega,0}$ if $U \gg t$:
\begin{align}
\Pq(\iw) \simeq \delta_{\omega,0} V^4 \chi_{\textrm{c},\bm{q}}/T,
\end{align}
where $\chi_{\textrm{c},\bm{q}}$ is the static susceptibility of the conduction electrons
defined by $\chi_{\textrm{c},\bm{q}}=T\sum_{\omega}[ X_{\textrm{c},\bm{q}} (\iw) - \bar{X}_{\textrm{c}} (\iw) ]$.
Replacing $\Pq(\iw)$ in Eq.~(\ref{eq:I_q}) and using $\phi(0)=4/U$ [from Eq.~(\ref{eq:phi-atom})], we obtain
\begin{align}
\Iq \simeq J_\textrm{K}^2 \chi_{\textrm{c},\bm{q}},
\end{align}
where the Kondo exchange coupling $J_\textrm{K}$ is defined by $J_\textrm{K}=4V^2/U$.
The above expression for $\Iq$ is consistent with the RKKY interaction.
\\

As demonstrated above, our SCL formula correctly reproduces the RKKY interaction in $f$ electron systems and the kinetic exchange interaction in $d$ electron systems, depending on the form of $\Pq(\iw)$ (delta function and constant, respectively).
It provides a solid ground for applications of the SCL formula.

\section{Extension to Multiorbital systems}
\label{sec:multiorbital}

We now recall the orbital indices in $\bchiloc$ and $\bXloc(\iw,\iw')$, and consider multiorbital systems (or spin-orbital coupled systems if the spin-orbit coupling exists).
In this case, we diagonalize $\bchiloc$ with respect to the indices $\alpha\equiv(1,2)$ and $\alpha'\equiv(3,4)$, and consider ``eigen-modes'' $\xi$ of local fluctuations as
\begin{align}
[\bchiloc]_{\alpha\alpha'}
= \sum_{\xi} W_{\alpha\xi} \chiloc^{\xi} W_{\xi\alpha'}^{\dag}.
\end{align}
Using the unitary matrix $\bm{W}$, we transform $\bXloc(\iw,\iw')$ into the eigen-mode basis
\begin{align}
\Xloc^{\xi}(\iw, \iw') \equiv
\sum_{\alpha\alpha'} W_{\xi\alpha}^{\dag} [\bXloc(\iw, \iw')]_{\alpha\alpha'} W_{\alpha'\xi}.
\label{eq:Xloc_eigen}
\end{align}
We neglect off-diagonal components ($\xi \neq \xi'$), since they disappear when summations over $\omega$ and $\omega'$ are taken.
Thus, the discussion up to here for the single-orbital model can be applied to each eigen-mode $\Xloc^{\xi}(\iw,\iw')$.

\begin{figure}[tb]
	\begin{center}
	\includegraphics[width=\linewidth]{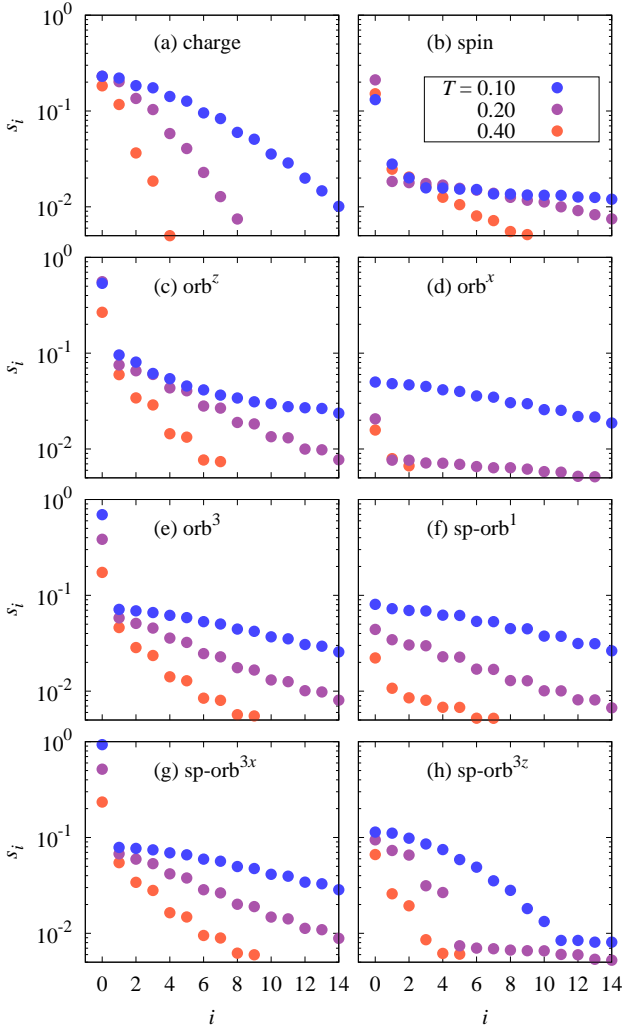}
	\end{center}
	\caption{Singular values $s_i^{\xi}$ of $\Xloc^{\xi}(\iw,\iw')$ in a two-orbital atomic model for different values of $T$. See Eq.~(\ref{eq:H-2orb}) for the explicit Hamiltonian ($\Delta/W=1.1$). The eight fluctuation modes in (a)--(h) are listed in Table~\ref{tab:op}.}
	\label{fig:SV-multiorb}
\end{figure}

As in Eq.~(\ref{eq:X-SVD}), we apply SVD to $\Xloc^{\xi}(\iw, \iw')$ and retain the leading term as
\begin{align}
\Xloc^{\xi}(\iw, \iw')
&= \sum_{i\geq0} s^{\xi}_i u^{\xi}_i(\iw) {v^{\xi}_i}^*(\iw')
\simeq s^{\xi}_0 u^{\xi}_0(\iw) {v^{\xi}_0}^*(\iw').
\label{eq:X-SVD-multiorb}
\end{align}
We show, in Fig.~\ref{fig:SV-multiorb}, an exemplary numerical result for the singular values $s_i^{\xi}$ in a two-orbital atom.
The explicit form of the interactions will be presented in Sec.~\ref{sec:results-multiorb}.
Here, we only emphasize that the eight independent fluctuation modes are classified into two types.
Figures~\ref{fig:SV-multiorb}(b), \ref{fig:SV-multiorb}(c), \ref{fig:SV-multiorb}(e), and \ref{fig:SV-multiorb}(g) exhibit a singular behavior at $s_0$, as in the spin channel in the Hubbard model [left panel in Fig.~\ref{fig:SV}(a)].
The singular enhancement of $s_0$ indicates the existence of local degrees of freedom, which gives rise to the Curie law in the local susceptibility.
For such fluctuation modes, the decoupling in Eq.~(\ref{eq:X-SVD-multiorb}) is justified.
On the other hand, the rest fluctuation modes [Figs.~\ref{fig:SV-multiorb}(a), \ref{fig:SV-multiorb}(d), \ref{fig:SV-multiorb}(f), and \ref{fig:SV-multiorb}(h)] are continuous around $i=0$ as in the charge channel in the Hubbard model [right panel in Fig.~\ref{fig:SV}(a)].
The decoupling is not valid for those modes.
However, since such fluctuations are absolutely small, the decoupling does not affect principal results in $\bchiq$.
Thus, the decoupling approximation in Eq.~(\ref{eq:X-SVD-multiorb}) is expected to provide a reasonable description in multiorbital models.

Using the decoupling in Eq.~(\ref{eq:X-SVD-multiorb}), we can represent Eq.~(\ref{eq:Xloc_eigen}) as follows:
\begin{align}
\bm{X}_\textrm{loc}(\iw,\iw') \simeq \bPhi(\iw) \bPhi(\iw'),
\label{eq:X-decouple-multiorb}
\end{align}
where
\begin{align}
[\bPhi(\iw)]_{\alpha\alpha'} &= \sum_{\xi} W_{\alpha\xi} \left( \sqrt{s_0^{\xi}} u_0^{\xi}(\iw) \right) W^{\dag}_{\xi\alpha'}.
\label{eq:Phi_L}
\end{align}
The phases of $u_0^{\xi}(\iw)$ and $v_0^{\xi}(\iw)$ have been fixed so that $u_0^{\xi}(\iw)={v_0^{\xi}}^*(\iw)$ (for details, see Appendix~\ref{app:u0-v0}).
The frequency dependence has thus been decoupled as in Eq.~(\ref{eq:X-decouple}), while the spin-orbital indices, $\alpha$ and $\alpha'$, are retained to be evaluated by the matrix product.

Once the decoupling in Eq.~(\ref{eq:X-decouple-multiorb}) is applied, the BS equation in Eq.~(\ref{eq:chiq-DMFT-2}) is evaluated to yield
\begin{align}
\bm{X}_{\bm{q}}(\iw,\iw') = \bPhi(\iw) [\bm{1} + \bm{\lambda}_{\bm{q}} + \bm{\lambda}_{\bm{q}}^2 + \cdots ] \bPhi(\iw'),
\label{eq:X-SCL-multiorb}
\end{align}
where we defined a multiorbital extension of $\lambda_{\bm{q}}$ in Eq.~(\ref{eq:lambda_q-1}) by
\begin{align}
\bm{\lambda}_{\bm{q}} = \sum_{\omega} \bPhi(\iw) \bPq(\iw) \bPhi(\iw).
\label{eq:lambda_q-multiorb}
\end{align}
We can now take the summations over $\omega$ and $\omega'$ independently in Eq.~(\ref{eq:X-SCL-multiorb}), and obtain the SCL formula for the spin and orbital susceptibility
\begin{align}
\bchiq^\textrm{SCL} &= \bchiloc^{1/2} (\bm{1} - \bm{\lambda}_{\bm{q}} )^{-1} \bchiloc^{1/2}.
\label{eq:chiq-SCL-lambda-multiorb}
\end{align}
This is the multiorbital extension of Eq.~(\ref{eq:chiq-SCL-lambda}).
Here, we used the relation $\overline{\bPhi} ^2 \simeq T \bchiloc$ with $\overline{\bPhi} \equiv T\sum_{\omega} \bPhi(\iw) e^{i\omega0^+}$ [multiorbital version of Eq.~(\ref{eq:s0-sumrule})].

As in the case of the single-orbital formula, we can represent $\bchiq^\textrm{SCL}$ in an RPA-like expression.
We define
\begin{align}
\bIq \equiv \bchiloc^{-1/2} \bm{\lambda}_{\bm{q}} \bchiloc^{-1/2},
\end{align}
and then rewrite Eq.~(\ref{eq:chiq-SCL-lambda-multiorb}) as
\begin{align}
\bchiq^\textrm{SCL} = (\bchiloc^{-1} - \bIq)^{-1}.
\label{eq:chiq-SCL-multiorb}
\end{align}
$\bIq$ represents the intersite spin and orbital interactions.
In the strong-coupling regime, we can eliminate $\bchiloc$ in the expression of $\bIq$.
Following the procedure leading to Eq.~(\ref{eq:I_q}), we can show
\begin{align}
&\bIq \simeq T\sum_{\omega} \bphi(\iw) \bPq(\iw) \bphi(\iw),
\label{eq:I_q-multiorb}
\end{align}
where
$\bphi(\iw) \equiv \bPhi(\iw) \overline{\bPhi}^{-1}$.
Since $\bphi(\iw)$ is normalized, $\bIq$ has no explicit dependence on local fluctuations, and reflects lattice properties predominantly.
We note again that $\bIq$ describes both the kinetic exchange process and the RKKY interactions depending on the functional form of $\bPq(\iw)$ as demonstrated in Sec.~\ref{sec:analytical}.

If the $\xi$-dependence in $\phi^{\xi}(\iw)$ is neglected as a rough approximation (see Sec.~\ref{sec:SCL3}), $\bIq$ is reduced to a simpler expression
\begin{align}
\bIq \simeq T\sum_{\omega} \bPq(\iw) \phi(\iw)^2,
\label{eq:I_q-multiorb-2}
\end{align}
which is similar to Eq.~(\ref{eq:I_q}).

\section{Approximation scheme}
\label{sec:approx}

Our susceptibility formulas [Eq.~(\ref{eq:chiq-SCL}) for single-orbital cases and Eq.~(\ref{eq:chiq-SCL-multiorb}) for multiorbital cases] simplify the process of computing $\bchiq$ in the DMFT.
In particular, it has been shown, in the single-orbital Hubbard model, that $T_\textrm{N}$ computed using the formula shows a complete agreement with the rigorous solution of the BS equation.
This result indicates that the decoupling of $\bXloc(\iw,\iw')$ essentially does not worsen the estimation of transition temperatures.
In a practical point of view, however, the computational cost is still not low enough, since the decoupling in Eq.~(\ref{eq:X-SVD}) requires as much information as in solving the BS equations.
Simpler calculations of $\bchiq$ are achieved by evaluating the decoupled functions $\bPhi(\iw)$ or $\bphi(\iw)$ \emph{without} SVD of $\Xloc^{\xi}(\iw,\iw')$.

\begin{figure}[tb]
	\begin{center}
	\includegraphics[width=\linewidth]{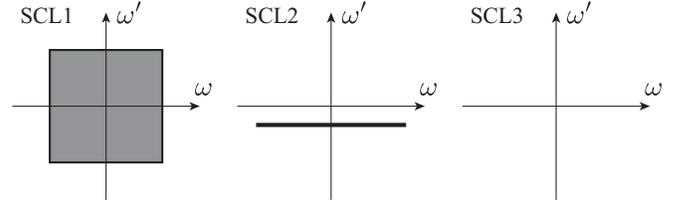}
	\end{center}
	\caption{Schematic figures showing how much information of $\Xloc(iw, iw')$ is necessary in each approximation scheme.}
	\label{fig:scl_scheme}
\end{figure}

In order to reduce the computational cost for $\bXloc(\iw,\iw')$,
we propose three levels of approximation in evaluating $\bPhi(\iw)$:
\begin{description}
\item[SCL1]
Using SVD [Eq.~(\ref{eq:X-SVD})]. Full information of $\bXloc(\iw, \iw')$ is required within a certain cutoff frequency. Results are already shown in Sec.~\ref{sec:numerical}.
\item[SCL2]
Estimating  $u_0^{\xi}(\iw)$ from one-dimensional data of $\bXloc(\iw, \iw')$. For example, $\iw$ is varied with fixing $\iw'=\iw_0$.
\item[SCL3]
Assuming an analytical form of $u^{\xi}_0(\iw)$ which is valid in the atomic limit. No calculation of $\bXloc(\iw, \iw')$ is required.
\end{description}
The difference in the necessary information of $\bXloc(\iw, \iw')$ is schematically depicted in Fig.~\ref{fig:scl_scheme}.
The cost of computing $\bXloc(\iw,\iw')$ is $\mathcal{O}(N_{\omega}^2)$, $\mathcal{O}(N_{\omega}^1)$, and $\mathcal{O}(1)$ for SCL1, SCL2, and SCL3, respectively.
Here, $N_{\omega}$ is the number of fermionic Matsubara frequencies in each axis.

\subsection{SCL2: A crude estimation of $u_0(\iw)$}
\label{sec:SCL2}
In the second scheme, SCL2, only a part of data in $\omega$-$\omega'$ plane is used.
This approximation is based on the fact that, if the decoupling in Eq.~(\ref{eq:X-decouple}) is exact, one-dimensional data (cut of the plane) is sufficient for evaluation of $u_0(\iw)$.
Actually, since the result depends on the location of the cut, one can improve accuracy by considering more than one cut and taking the average over different estimations.
Technical details are presented in Appendix~\ref{app:scl2} for a practical implementation.

\begin{figure}[tb]
	\begin{center}
	\includegraphics[width=\linewidth]{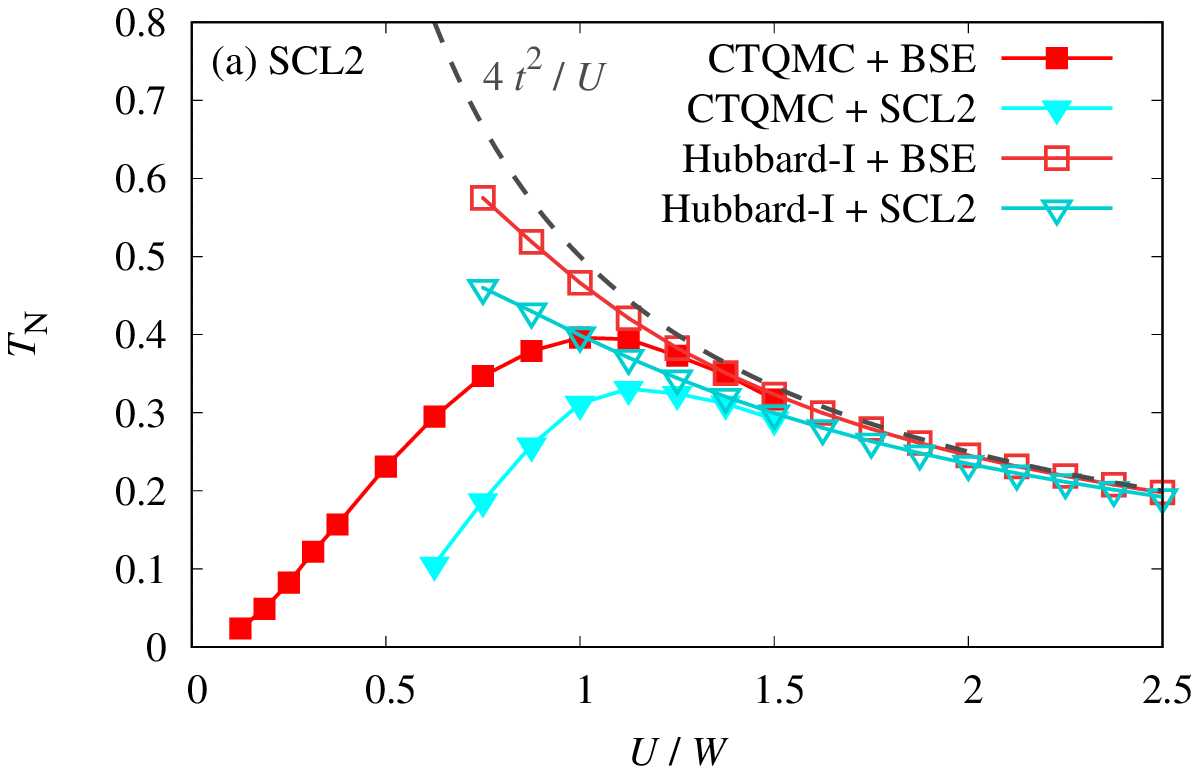}
	\includegraphics[width=\linewidth]{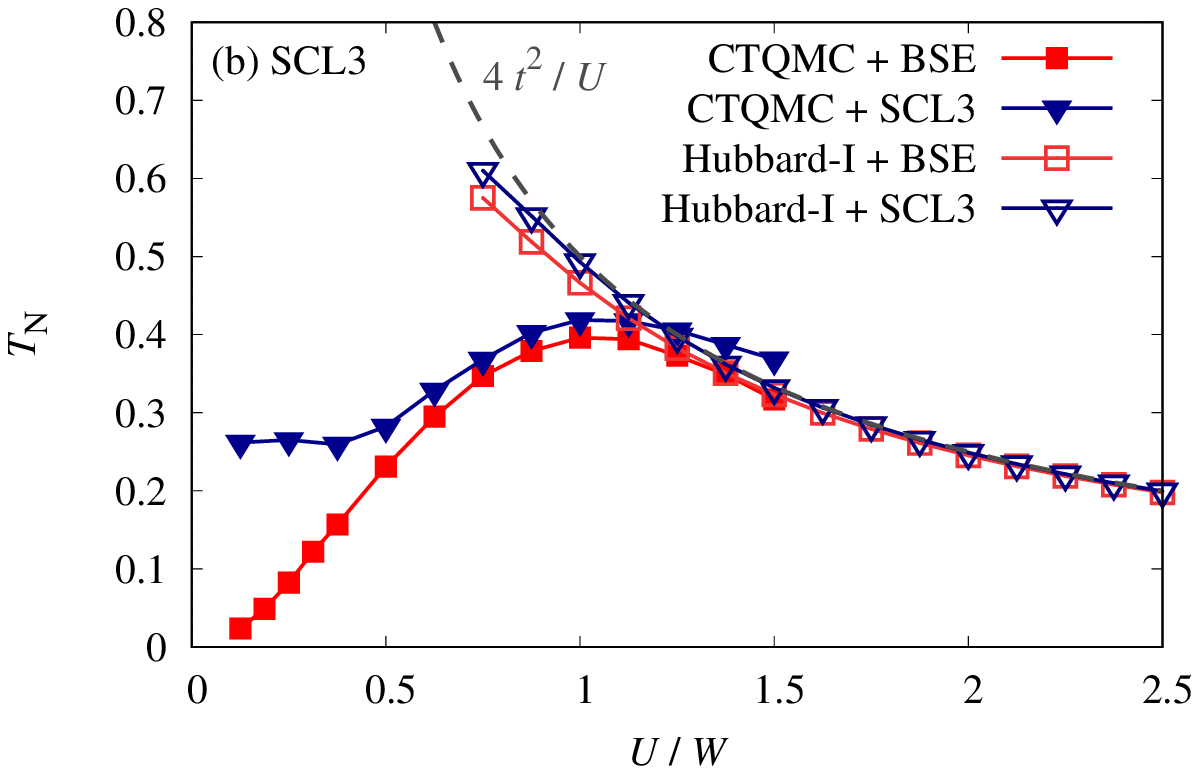}
	\end{center}
	\caption{Comparison between $T_\textrm{N}$ computed by an approximation scheme and that computed by the original BS equation. (a) SCL2 scheme, and  (b) SCL3 scheme.}
	\label{fig:scl_approx}
\end{figure}

The phase diagram obtained by the SCL2 scheme is shown in Fig.~\ref{fig:scl_approx}(a).
The value of $T_\textrm{N}$ turns out to be underestimated, in particular, in the weak-coupling regime, $U/W\lesssim 1.0$.
The deviation becomes smaller as $U$ increases, and
the SCL2 reproduces $T_\textrm{N}$ almost perfectly in the strong-coupling regime, $U/W \gtrsim 1.5$.

The fact that the SCL2 scheme underestimates $T_\textrm{N}$ can be explained in terms of a variational principle as follows.
We first remark that the square of the singular value, $s_i^2$, corresponds to the eigenvalue of $\bXloc \bXloc^{\dag}$.
Hence, the decoupling approximation, Eq.~(\ref{eq:X-SVD}), is represented by
$s_0^2 \simeq \langle u_0 | \bXloc \bXloc^{\dag} | u_0 \rangle$.
The right-hand side becomes maximum when $u_0(\iw)$ is the true eigenvector of $\bXloc \bXloc^{\dag}$.
In other words, an approximate evaluation of $u_0(\iw)$ results in a smaller estimation of $s_0$, namely, the underestimate of $\chiq$.

This fact gives us ``controlability" in applications of the SCL2 scheme.
Since the SCL2 scheme underestimates all fluctuation modes, a competition among different modes would still be captured correctly.
Indeed, the result for the two-orbital model in Sec \ref{sec:results-multiorb} supports the validity of the SCL2 approximation.

\subsection{SCL3: Two-pole approximation}
\label{sec:SCL3}

In the single-orbital model, the function $u_0(\iw)$ is reduced to a simple two-pole function (Lorentzian) in the atomic limit, as discussed in Sec.~\ref{sec:u0}.
This results from the fact that the intermediate empty and doubly occupied states are both singlet.
In multiorbital models, on the other hand, $u_0(\iw)$ is not of Lorentzian, since intermediate states have multiplet.
Nevertheless, the two-pole function gives a reasonable approximation for relevant fluctuation modes that follow the Curie law.
Detailed analysis in a two-orbital model is presented in Appendix~\ref{app:scl3}.
For this reason, we adopt the two-pole function also in multiorbital cases and construct an approximation that does not require explicit computations of $\bm{X}_\textrm{loc}(\iw,\iw')$.

We generalize the two-pole function in Eq.~(\ref{eq:phi-atom}) into multiorbital cases as\footnote{This function fulfills the normalization condition $\overline{\phi^{\xi}}=1$ only at low-$T$ of $T \ll E_{\pm}^{\xi}$. However, it does not affect an estimation of transition temperatures. Furthermore, the weights of the two terms can actually be different from each other (keeping the sum at 2), but we do not introduce such a parameter for simplicity.}
\begin{align}
\phi^{\xi}(\iw) = \frac{1}{\iw+E_{-}^{\xi}} - \frac{1}{\iw-E_{+}^{\xi}},
\label{eq:phi-gen}
\end{align}
where
$E_{-}^{\xi} = E_{n-1} - E_{n} >0$ and
$E_{+}^{\xi} = E_{n+1} - E_{n} >0$.
Here, $E_{n}$ denotes a ``typical'' value of $n$-particle states, e.g., the average or the lowest energy.
In general, $E_{\pm}^{\xi}$ depends on $\xi$, because the dominant contribution to $\Xloc^\xi(\iw,\iw')$ occurs from different $(n\pm1)$-particle states depending on $\xi$.
Once the parameters $E_{\pm}^{\xi}$ are fixed, we can readily evaluate $\bchiq^\textrm{SCL}$ using Eqs.~(\ref{eq:chiq-SCL-multiorb}) and (\ref{eq:I_q-multiorb}), which we refer to as SCL3 approximation.
In practical applications, it is convenient to replace $E_{\pm}^{\xi}$ with the lowest excitation energy $E_{\pm,\textrm{min}}$ estimated in the local Hamiltonian.
For details, see the example in Appendix~\ref{app:scl3} and Sec.~\ref{sec:results-multiorb}.

The accuracy of the SCL3 scheme is examined in Fig.~\ref{fig:scl_approx}(b) for the single-orbital model,
in which $E_{\pm}^{\xi}$ can be uniquely fixed at $E_{\pm}^{\xi}=U/2$ for the half-filling.
The transition temperature turns out to be slightly overestimated for $U/W \gtrsim 0.5$ and converges to the strong-coupling tail.
The deviation is large in the weak-coupling regime of $U/W \lesssim 0.5$, since the analytic expression in Eq.~(\ref{eq:phi-gen}) is justified only for large-$U$.

Though the SCL3 approximation gives reasonable results for the single-orbital model, the situation is rather complicated in multiorbital models.
The SCL3 scheme corresponds to an approximation which neglects splitting of energy levels of excited multiplet, and may result in a misjudgment of competing orders.
See the next section for a test in a multiorbital model.

\section{Application: Two-orbital model}
\label{sec:results-multiorb}

\subsection{Model and background}

For a demonstration of the multiorbital version of the strong-coupling formula, Eq.~(\ref{eq:chiq-SCL-multiorb}), 
we consider a two-orbital model with an energy splitting $\Delta$.
Letting the spin index $\sigma=\uparrow$, $\downarrow$, and the orbital index $\tau=\textrm{a}$, b, the Hamiltonian reads
\begin{align}
\mathcal{H} &= \sum_{\bm{k}\tau\sigma} \epsilon_{\bm{k}} c_{\tau\sigma}^{\dag} c_{\tau\sigma}
+ \frac{\Delta}{2} \sum_{i\sigma} [n_{i\textrm{b}\sigma} - n_{i\textrm{a}\sigma} ]
+ U \sum_{i\tau} n_{i\tau\uparrow} n_{i\tau\downarrow}
\nonumber \\
&+ \sum_{i, \tau>\tau', \sigma} [ U' n_{i\tau\sigma} n_{i\tau' \overline{\sigma}}
+ (U'-J) n_{i\tau\sigma} n_{i\tau'\sigma} ]
\nonumber \\
&+ J \sum_{i,\tau \neq \tau'}
 ( c_{i\tau\downarrow}^{\dag} c_{i\tau'\uparrow}^{\dag} c_{i\tau'\uparrow} c_{i\tau\downarrow}
+  c_{i\tau\uparrow}^{\dag} c_{i\tau\downarrow}^{\dag} c_{i\tau'\uparrow} c_{i\tau'\downarrow}).
\label{eq:H-2orb}
\end{align}
The kinetic energy term is assumed to be orbital-diagonal.
We consider the tight-binding band on the square lattice in common with the previous example in Eq.~(\ref{eq:H-Hubbard}).
The second term expresses the energy splitting $\Delta$ between the two bands.
The rest terms are the Slater-Kanamori interactions.
We consider the half-filled case, $n=2$, and vary $\Delta$ keeping other parameters, $U$, $U'$, and $J$.
We fix $U$ at $U/W=1.5$, and use the Hubbard-I solver, which is expected to be reasonable at this parameter as demonstrated in the case with the single-band model (see for example Fig.~\ref{fig:phase_rrpa}).
The Hund's coupling $J$ is fixed at $J=U/4$, and $U'$ at $U'=U-2J$ to keep the rotational symmetry of the orbitals when $\Delta=0$.

\begin{figure}[tb]
	\begin{center}
	\includegraphics[width=0.95\linewidth]{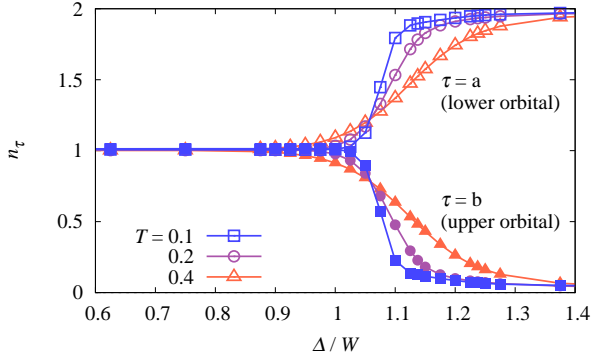}
	\end{center}
	\caption{The orbital-dependent occupation number $n_{\tau}$ in the two-orbital model.}
	\label{fig:2orb-occup}
\end{figure}

This model has been discussed in the context of the spin-state transition~\cite{Werner07,Suzuki09,Kunes11b} and an excitonic insulator (EI)~\cite{Kaneko12,Kunes14a,Kunes14b,Nasu16,Hoshino16}.
Let us first review the basic properties of this model.
When $\Delta \ll W$, the two orbitals are almost degenerate, and the spin-triplet state is favored according to the Hund's coupling.
The two orbitals are equally occupied as shown in Fig.~\ref{fig:2orb-occup}.
When $\Delta \gg W$, on the other hand, only the lower band is occupied and the spin state becomes singlet.
It is known that the change between the two spin-states becomes more abrupt as $T$ is lowered~\cite{Werner07} as seen around $\Delta \simeq W$ in Fig.~\ref{fig:2orb-occup}.
Hence, large orbital fluctuations are expected.
As a results of competing spin and orbital fluctuations, a spin-orbital coupled state called EI state emerges in this region~\cite{Kaneko12,Kunes14a}.
We examine, in the following, whether this non-trivial ordered state can be reproduced by our simplified calculation scheme.

\subsection{Susceptibilities}

\begin{table}[tb]
	\caption{Classification of 16 operators $\sigma^{\xi}\tau^{\eta}$ in the two-orbital system. The column `irrep' shows irreducible representation in cubic point group in Mulliken and Bethe notation. The index g and u (or $+$ and $-$) represent time-reversal even and odd, respectively.}
	\label{tab:op}
	\begin{tabular}{lcccc}
	\hline
	label & operator(s) & degeneracy & \multicolumn{2}{c}{irrep in O$_h$} \\
	\hline
	charge & $\sigma^0 \tau^0$ & 1 & A$_\textrm{1g}$ & $\Gamma_1^+$ \\
	spin (AFM) & $\sigma^x \tau^0$, $\sigma^y \tau^0$ $\sigma^z \tau^0$ & 3 & T$_\textrm{1u}$ & $\Gamma_4^-$ \\
	orb$^z$ (HS/LS) & $\sigma^0 \tau^z$ & 1 & \multirow{2}{*}{$\Big\}$ E$_\textrm{g}$ \ \ } & \multirow{2}{*}{$\Gamma_3^+$} \\
	orb$^x$ & $\sigma^0 \tau^x$ & 1 & & \\
	orb$^\textrm{3}$ & $\sigma^x \tau^y$, $\sigma^y \tau^y$, $\sigma^z \tau^y$ & 3 & T$_\textrm{2g}$ & $\Gamma_5^+$ \\
	sp-orb$^1$ & $\sigma^0 \tau^y$ & 1 & A$_\textrm{2u}$ & $\Gamma_2^-$ \\
	sp-orb$^{3x}$ (EI) & $\sigma^x \tau^x$, $\sigma^y \tau^x$, $\sigma^z \tau^x$ & 3 & T$_\textrm{1u}$ & $\Gamma_4^-$ \\
	sp-orb$^{3z}$ & $\sigma^x \tau^z$, $\sigma^y \tau^z$, $\sigma^z \tau^z$ & 3 & T$_\textrm{2u}$ & $\Gamma_5^-$ \\
	\hline
	\end{tabular}
\end{table}

There are 16 fluctuation modes represented by the operator $O_{\tau\sigma, \tau'\sigma'}$ defined in Eq.~(\ref{eq:O}).
It is convenient to represent the spin and orbital indices in terms of the Pauli matrix $\sigma^{\xi}$ ($\xi=0,x,y,z$) as
\begin{align}
O^{\xi\eta} = \sum_{\sigma\sigma'\tau\tau'} c_{\tau\sigma}^{\dag} \sigma^{\xi}_{\sigma\sigma'} \sigma^{\eta}_{\tau\tau'} c_{\tau'\sigma'}
\equiv \sigma^{\xi}\tau^{\eta}.
\end{align}
These operators are classified into eight classes as summarized in Table~\ref{tab:op}.
The number of degeneracy is either 1 or 3, where threefold degeneracy is due to the spin part, ($\sigma^x$, $\sigma^y$, $\sigma^z$), while no extra degeneracy takes place because the orbital indices are all independent.
For reference, correspondence to the irreducible representations in the cubic point-group symmetry is also listed~\cite{Ohkawa83,Shiina97}.
The indices g (u) and $+$ ($-$) stand for time-reversal even (odd).
Here, the odd time-reversal symmetry comes from $\sigma^x$, $\sigma^y$, $\sigma^z$, and $\tau^y$.
The third to fifth classes (labeled with orb) are kinds of orbital orders, and the 6th to 8th (labeled with sp-orb) are spin-orbital coupled orders with broken time-reversal symmetry.
In particular, the 3rd operator (orb$^z$) is the same as the level splitting, and its staggard order corresponds to high-spin--low-spin (HS/LS) order.
The 7th operator (sp-orb$^{3x}$) corresponds to the EI ordered state.

\begin{figure*}[tbp]
	\begin{center}
	\includegraphics[width=0.95\linewidth]{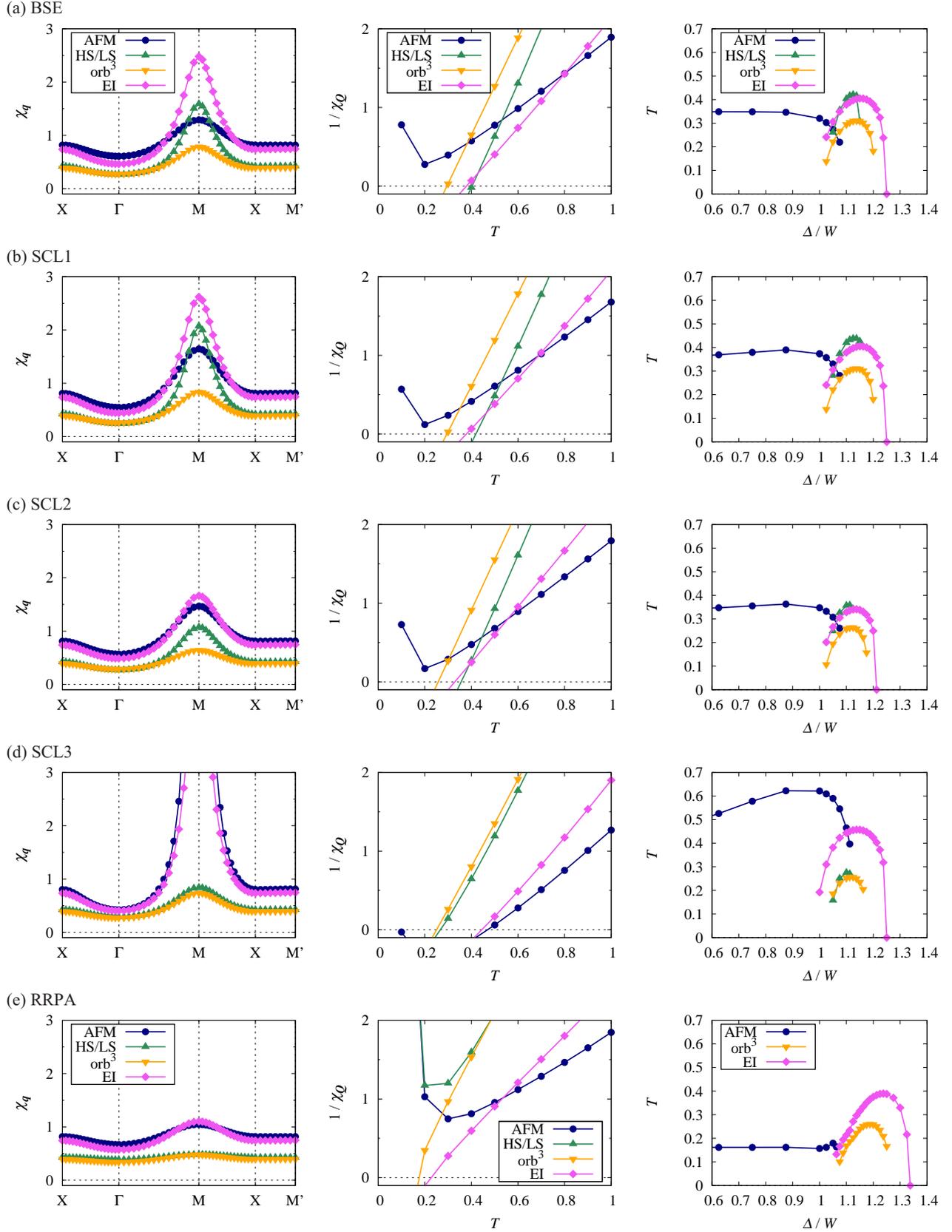}
	\end{center}
	\caption{Comparison of different approximation schemes in the two-orbital Hubbard model: (a) the original BS equation, (b) SCL1, (c) SCL2, (d) SCL3, and (e) RRPA. In each row, (Left) $\bm{q}$-dependence of $\chiq$ for fixed $\Delta/W=1.1$ and $T=0.5$, For the symbols of $\bm{q}$-points, see the caption of Fig.~\ref{fig:chiq_path}. (Center) $T$-dependence of $1/\chi_{\bm{Q}}$ for $\Delta/W=1.1$, and (Right) transition temperatures as a function of $\Delta/W$. Only 4 relevant fluctuation modes out of 8 are plotted. }
	\label{fig:2orb}
\end{figure*}

The left panel of Fig.~\ref{fig:2orb}(a) shows the $\bm{q}$-dependence of $\chiq$ obtained by solving the original BS equation in Eq.~(\ref{eq:chiq-DMFT}).
We chose the parameter $\Delta/W=1.1$, where multiple fluctuation modes are competing.
It turns out that $\bchiq$ exhibits enhancement at M-point [$\bm{q}=(\pi,\pi)\equiv \bm{Q}$].
The leading fluctuations are the EI order, HS/LS order, AFM order, and threefold orbital fluctuations (orb$^3$).
Other fluctuation modes exhibit no distinguished enhancement (not shown in the figure).
The transition temperatures are determined from the divergence of $\chi_{\bm{Q}}$ [the center panel of Fig.~\ref{fig:2orb}(a)].
The obtained phase diagram [the right panel of Fig.~\ref{fig:2orb}(a)] includes AFM phase for $\Delta/W \lesssim 1.0$, and EI and HS/LS phases for $\Delta/W \gtrsim 1.0$.
The HS/LS order appears only in the limited region at finite-$T$ because this state has degeneracy without additional symmetry breaking.
The orb$^3$ order is not realized since other transition takes place at higher $T$.

Let us now examine the accuracy of other approximation schemes.
Figure~\ref{fig:2orb}(b) shows the SCL1 result. 
It is quite similar to Fig.~\ref{fig:2orb}(a) with only a little enhancement of the transition temperatures, 
justifying the validity of our SCL formula even for the multi-orbital systems.

The SCL2 result in Fig.~\ref{fig:2orb}(c)
also reproduces the phase diagram very well. 
The qualitative feature such as the severe competition between EI and HS/LS orders are correctly reproduced. 
Quantitatively, the transition temperatures are slightly underestimated, in consistent with the conclusion in Sec.~\ref{sec:SCL2}.

The result in the SCL3 scheme is shown in Fig.~\ref{fig:2orb}(d).
In this calculation, we neglected the $\xi$-dependence of $\phi^{\xi}(\iw)$ for simplicity, and employed Eq.~(\ref{eq:I_q-multiorb-2}).
The parameter $E_{\pm}^{\xi}$ in Eq.~(\ref{eq:phi-gen}) were replaced with the lowest excitation energy $E_{\pm, \textrm{min}}$ determined from the eigenvalues of the local Hamiltonian.
Despite these approximations, we see the diverging susceptibility for the four enhanced modes.
However, the estimate of the transition temperature is less accurate than those by SCL1 and SCL2.
The phase diagram shows that the AFM fluctuations are largely overestimated, while the fluctuations for the HS/LS order is underestimated so that the HS/LS order is missing (covered by the EI order).
The source of these artifacts is discussed in the next subsection in terms of the effective nonlocal interactions.

Finally, Fig.~\ref{fig:2orb}(e) shows the RRPA result.
Although the RRPA reproduces the divergence of susceptibility in AFM, EI, and orb$^3$ orders, the HS/LS  susceptibility are largely underestimated and does not diverge at finite temperature. 
This artifact suggests that, as in the single-orbital case, the usage of the RRPA in the strong-coupling regime is not justified. 

\subsection{Effective nonlocal interactions}

\begin{figure}[tb]
	\begin{center}
	\includegraphics[width=0.70\linewidth]{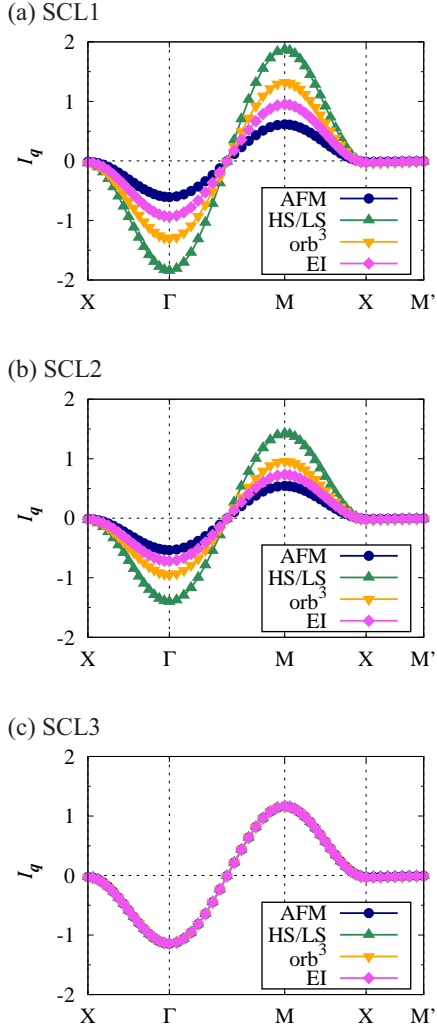}
	\end{center}
	\caption{The $\bm{q}$-dependence of $\Iq$ for fixed $\Delta/W=1.1$ and $T=0.5$. (a) SCL1, (b) SCL2, (c) SCL3.}
	\label{fig:2orb-Iq}
\end{figure}

In our SCL formula, it is possible to derive the effective nonlocal interactions $\Iq$, which enters in the RPA-like susceptibility formula in Eq.~(\ref{eq:chiq-SCL-multiorb}).
Figure~\ref{fig:2orb-Iq} shows $\Iq$ computed in the three approximate schemes.
For all schemes, the momentum dependence exhibits the simple cosine form, $\cos k_x + \cos k_y$, as in the strong-coupling regime of the single-band Hubbard model ($U/W=1.5$ in Fig.~\ref{fig:Iq_path}).
It indicates that the effective interaction works only between nearest-neighbor sites.

With the results of $\Iq$ in Fig.~\ref{fig:2orb-Iq}, we can understand the results of each SCL scheme in Fig.~\ref{fig:2orb}.
For SCL1 and SCL2, the HS/LS mode has the strongest interaction, which helps the realization of the HS/LS order around $\Delta/W=1.1$.
On the other hand, the SCL3 scheme exhibits no mode-dependence in $\Iq$, and hence underestimates the HS/LS transition temperature in Fig.~\ref{fig:2orb}(d).

The artificial degeneracy of $\Iq$ in the SCL3 scheme originates from the approximation that we neglected the $\xi$-dependence of $E_{\pm}^{\xi}$ (instead $E_{\pm, \textrm{min}}$ is used for all modes) in Eq.~(\ref{eq:phi-gen}).
According to the SVD analysis in Appendix~\ref{app:scl3}, $E_{\pm}^\textrm{HS/LS}$ is actually smaller than the others, so replacing $E_{\pm}^\textrm{HS/LS}$ with $E_{\pm, \textrm{min}}$ reduces the HS/LS fluctuation.
On the contrary, the actual value of $E_{\pm}^\textrm{spin}$ is about two times larger than $E_{\pm, \textrm{min}}$, meaning that the use of $E_{\pm, \textrm{min}}$ results in an overestimate of the spin exchange interactions.
The above analysis of $\Iq$ indicates that a proper estimation of $E_{\pm}^{\xi}$ is needed when the SCL3 scheme is applied to competing orders.

\section{Summary}
\label{sec:summary}
The BS equation for the calculations of $\bchiq$ in the DMFT was reduced to an RPA-like form called the SCL formula [Eq.~(\ref{eq:chiq-SCL}) for single-orbital cases and Eq.~(\ref{eq:chiq-SCL-multiorb}) for multiorbital cases].
Its derivation is based on the fact that the two-particle Green function $\bXloc(\iw, \iw')$ can be decoupled [Eqs.~(\ref{eq:X-decouple}) and (\ref{eq:X-decouple-multiorb})] if local moments are well-defined and there exist local fluctuations which give rise to the Curie law.
Since a long-range order often occurs in such situations, the SCL formula is suitable for the use of investigations of phase transitions.
Indeed, the SCL formula is found to yield an accurate transition temperature over a surprisingly wide range of parameters, though $\bchiq$ is overestimated at higher temperatures in the weak-coupling region.

So far, the bottleneck in computing $\bchiq$ was twofold: (i) calculations of $\bXloc(\iw,\iw')$ in the effective local problem and (ii) solving the BS equation with large matrices.
The SCL formula provides the correct phase diagram in accuracy comparable to the BS equation [see SCL1 results in Figs.~\ref{fig:phase_scl} and \ref{fig:2orb}(b)], and therefore settles the latter issue.
Furthermore, approximate estimations of the function $\bPhi(\iw)$ or $\bphi(\iw)$ (SCL2 and SCL3 schemes) relieve the former issue.
The cost of computing $\bXloc(\iw,\iw')$ is reduced from $\mathcal{O}(N_{\omega}^2)$ to $\mathcal{O}(N_{\omega}^1)$ in the SCL2 scheme and $\mathcal{O}(1)$ in the SCL3 scheme, where $N_{\omega}$ is the number of fermionic Matsubara frequencies.
We note that, in the SCL3 scheme, the mode-dependence of the effective excitation energy, $E_{\pm}^{\xi}$, is relevant when different types of fluctuation modes are competing.

By combining the three levels of approximation in the SCL formula and the RRPA formula in the weak-coupling side, one can evaluate $\bchiq$ in the DMFT without solving the BS equation.
It will enable systematic investigations of complicated spin-orbital orders in multiorbital systems, in particular, realistic $d$ and $f$-electron materials within the DFT+DMFT framework.

\begin{acknowledgments}
We acknowledge useful discussions with S. Ishihara, M. Naka, J. Kune\ifmmode \check{s}\else \v{s}\fi{}, H. Ikeda, Y. \=Ono, T. Yamada, Y. Iizuka.
The numerical calculations were performed using the following open source libraries and softwares:
{\tt TRIQS}\cite{triqs},
{\tt DFTTools}\cite{dfttools},
{\tt pomerol}\cite{pomerol},
{\tt pomerol2triqs}, and
{\tt pomerol2triqs1.4}.
We were supported by JSPS KAKENHI Grant No. 18H01158.
J.O. was supported by JSPS KAKENHI Grant No. 18H04301 (J-Physics).
K.Y. was supported by Building of Consortia for the Development of Human Resources in Science and Technology, MEXT, Japan.
H.S. was supported by JSPS KAKENHI Grant No. 16K17735.
Y.N. was supported by JSPS KAKENHI Grant Nos. 16H06345 and 17K14336.
\end{acknowledgments}

\appendix

\section{Hubbard-I approximation}
\label{app:Hubbard-I}

In practical computations for multiorbital models with complicated interactions, it is difficult to solve the effective impurity problem at arbitrary temperature and interaction strength without approximation.
Furthermore, the self-consistency loop of the DMFT may become unstable in the strong-coupling regime.
In this situation, it is useful to consider the atomic limit expressed by
\begin{align}
\Delta(i\omega)=0,
\label{eq:atomic_limit}
\end{align}
and plug only the one-body levels into the impurity problem.
This is called the Hubbard-I approximation in the context of Mott gap in the single-particle excitation spectra.

With the approximation in Eq.~(\ref{eq:atomic_limit}), we can compute the local self-energy $\Sigma_{\rm atom}(\iw)$ and the two-particle Green function $X_\textrm{atom}(\iw, \iw'; \iW)$ by diagonalizing the atomic Hamiltonian explicitly.
We used {\tt pomerol} implementation\cite{pomerol} in our calculations.

Some remarks are made on the atomic-limit calculations for two-particle quantities.
Because of the absence of self-consistent determination of $\Delta(\iw)$, the self-consistency condition of DMFT
\begin{align}
G_\textrm{imp}(\iw) = \frac{1}{N} \sum_{\bm{k}} G(\bm{k}, \iw) \equiv G_{\rm loc}(\iw),
\end{align}
is {\em not} fulfilled.
Here, $G_\textrm{imp}(\iw)$ denotes the Green function computed in the impurity problem.
It leads that a similar equality for the bare two-particle Green function
\begin{align}
\bm{X}_{0, \textrm{imp}}(\iW) = \frac{1}{N} \sum_{\bm{q}} \bm{X}_0(\bm{q}, \iW) \equiv \bm{X}_{0,\textrm{loc}}(\iW),
\end{align}
is also violated, though the self-consistently determined solution of the DMFT does have this equality.
Here, $\bm{X}_{0, \textrm{imp}}(\iW)$ is given by $[\bm{X}_{0, \textrm{imp}}]_{\omega,\omega'}(\iW)=-\delta_{\omega\omega'} G_\textrm{imp}(\iw) G_\textrm{imp}(\iw+\iW)$.
Thus, there is a choice between $\bm{X}_{0, \textrm{imp}}(\iW)$ and $\bm{X}_{0,\textrm{loc}}(\iW)$ in solving the BS equation in Eq.~(\ref{eq:chiq-DMFT}).
We found that using $\bm{X}_{\rm 0,loc}(\iW)$ gives reasonable results, because the difference between $\bm{X}_0(\bm{q},\iW)$ and $\bm{X}_{\rm 0,loc}(\iW)$ plays an essential role in the $\bm{q}$-dependence of the susceptibility.

\section{Frequency cutoff}
\label{app:freq_cutoff}
In practical computations of the BS equation, one needs to introduce cutoff in the fermionic frequencies.
Here, we present a practical expression that is suitable for introducing cutoff.
We begin with the BS equation in the form of Eq.~(\ref{eq:chiq-DMFT-2}),
which is rewritten as follows:
\begin{align}
\bXq &= \bXloc + \Delta\bXq, \\
\Delta\bXq &= \bXloc ( \bPq^{-1} - \bXloc )^{-1} \bXloc.
\label{eq:ZQ2}
\end{align}
Here, $\Delta\bXq$ describes the deviation with respect to $\bXloc$.
Taking the summations over the fermionic Matsubara frequencies, we obtain
\begin{align}
\bchiq &= \bchiloc + T \sum_{\omega,\omega'} \Delta\bXq.
\end{align}
The infinite summations in the first term have been taken exactly.
We now introduce a frequency cutoff for the summations in the second term to evaluate the deviation from $\bchiloc$.

\section{Relations between $u_0(i\omega)$ and $v_0(i\omega)$}
\label{app:u0-v0}

In this section, we derive relations between $u_0(i\omega)$ and $v_0(i\omega)$ from symmetry properties of two-particle Green functions.
We first use the relation
\begin{align}
\Xloc(\iw,\iw') &= \Xloc(\iw',\iw),
\label{eq:X-sym1}
\end{align}
which implies that $u_0(\iw)$ and $v_0(\iw)$ are basically equivalent.
Replacing $\Xloc(\iw,\iw')$ with the decoupling formula in Eq.~(\ref{eq:X-SVD}), we can prove $u_0(\iw)=C v_0^*(\iw)$, where $C$ is a constant.
Furthermore, the normalization condition for $u_0(\iw)$ and $v_0(\iw)$ leads to $|C|=1$, and thus the relation is expressed as
\begin{align}
u_0(\iw) = e^{i\theta} v_0^*(\iw),
\label{eq:u0-v0-conj}
\end{align}
where $\theta$ is a real number.

Next, we use another relation that connects $\iw$ and $-\iw$
\begin{align}
\Xloc^*(\iw,\iw') &= \Xloc(-\iw,-\iw').
\label{eq:X-sym2}
\end{align}
The same procedure as above leads to a similar relation between $u_0(\iw)$ and $u_0^*(-\iw)$, namely,
$u_0(\iw) = e^{i\theta'} u_0^*(-\iw)$.
Furthermore, combining with Eq.~(\ref{eq:u0-v0-conj}), we can prove $\theta=\theta'$, and thus obtain
\begin{align}
u_0(\iw) &= e^{i\theta} u_0^*(-\iw),
\label{eq:u0-conj}
\\
v_0(\iw) &= e^{i\theta} v_0^*(-\iw).
\label{eq:v0-conj}
\end{align}
It is also possible to eliminate the phase factor by combining Eqs.~(\ref{eq:u0-v0-conj}) and (\ref{eq:u0-conj}) to yield
\begin{align}
u_0(\iw) = v_0(-\iw),
\label{eq:u0-v0}
\end{align}
which does not depend on the choice of $\theta$.

In practical calculations, it is convenient to fix the phase so that $\theta=0$.
To this end, we define
$\widetilde{u}_0(\iw) \equiv u_0(\iw) e^{-\theta/2}$ and
$\widetilde{v}_0(\iw) \equiv v_0(\iw) e^{-\theta/2}$,
and thus $\widetilde{u}_0(\iw)$ and $\widetilde{v}_0(\iw)$ follow the above equations with $\theta=0$, namely,
\begin{align}
\widetilde{u}_0(\iw) &= \widetilde{v}_0^*(\iw),\\
\widetilde{u}_0(\iw) &= \widetilde{u}_0^*(-\iw).
\end{align}

\section{Derivation of $\Lambda_{\vect{q}}(i\omega)$ in the Hubbard model}
\label{app:Pq-Hubbard}

We derive the analytical expression for $\Pq(i\omega)$ in Eq.~(\ref{eq:Pq-Hubbard}).
The single-particle Green function in the Hubbard model is given by
\begin{align}
G_{\bm{k}}(\iw) = \frac{1}{\iw + \mu - \epsilon_{\bm{k}} - \Sigma(\iw)}.
\end{align}
In the atomic limit, the local Green function $G_\textrm{loc}(\iw)$ is 
\begin{align}
G_\textrm{loc}(\iw) \simeq \frac{1}{\iw + \mu - \Sigma(\iw)}.
\end{align}
We expand $G_{\bm{k}}(\iw)$ with respect to $\epsilon_{\bm{k}}$ and retain the first correction
\begin{align}
G_{\bm{k}}(\iw) \simeq G_\textrm{loc}(\iw) + G_\textrm{loc}(\iw) \epsilon_{\bm{k}} G_\textrm{loc}(\iw).
\end{align}
Using this expression, $\Xq^0(\iw; \iW)$ is evaluated as
\begin{align}
\Xq^0(\iw; \iW) \simeq \Xloc^0(\iw; \iW)
- \Xloc^0(\iw; \iW)^2 \langle \epsilon_{\bm{k}} \epsilon_{\bm{k}+\bm{q}} \rangle_{\bm{k}}.
\end{align}
Here, we used $\langle \epsilon_{\bm{k}} \rangle_{\bm{k}}=0$.
Inserting this approximated result into $\Pq(\iw; \iW)$ in Eq.~(\ref{eq:P_q}), we obtain
\begin{align}
\Pq(\iw; \iW) \simeq -\langle \epsilon_{\bm{k}} \epsilon_{\bm{k}+\bm{q}} \rangle_{\bm{k}}.
\end{align}
Taking the average over $\bm{k}$ and using $\langle \gamma_{\bm{k}} \gamma_{\bm{k}+\bm{q}} \rangle_{\bm{k}} =\gamma_{\bm{k}}/2$ for a nearest-neighbor hopping model, we obtain Eq.~(\ref{eq:Pq-Hubbard}).

\section{Derivation of $\Lambda_{\vect{q}}(i\omega)$ in the periodic Anderson model}
\label{app:Pq-PAM}

The analytical expression for $\Pq(i\omega)$ in Eq.~(\ref{eq:Pq-PAM}) is derived in this appendix.
The single-particle Green function in the periodic Anderson model is given by
\begin{align}
G_{\bm{k}}(\iw) = \frac{1}{\iw - \epsilon_f - V_{\bm{k}}^2 g_{\textrm{c},\bm{k}}(\iw) - \Sigma(\iw)},
\end{align}
where $g_{\textrm{c},\bm{k}}(\iw)$ is the conduction electron Green function defined by $g_{\textrm{c},\bm{k}}(\iw)=1/(\iw+\mu-\epsilon_{\bm{k}})$.
The local Green function $G_\textrm{loc}$ is given by
\begin{align}
G_\textrm{loc}(\iw) = \frac{1}{\iw - \epsilon_f - \Delta(\iw) - \Sigma(\iw)}.
\end{align}
In the atomic limit, $\Delta(\iw)$ is approximated into $\Delta(\iw) \simeq \langle V_{\bm{k}}^2 g_{\textrm{c},\bm{k}}(\iw) \rangle_{\bm{k}}$, and 
we expand
$G_{\bm{k}}^{-1}(\iw)=G_\textrm{loc}^{-1}(\iw) - [V_{\bm{k}}^2 \epsilon_{\bm{k}} - \Delta(\iw)]$
around $G_\textrm{loc}(\iw)$ as
\begin{align}
G_{\bm{k}}(\iw) \simeq G_\textrm{loc}(\iw) + G_\textrm{loc}(\iw) [V_{\bm{k}}^2 \epsilon_{\bm{k}} - \Delta(\iw)] G_\textrm{loc}(\iw).
\end{align}
Then, $\Xq^0(\iw; \iW)$ is evaluated as
\begin{align}
&\Xq^0(\iw; \iW) \simeq \Xloc^0(\iw; \iW)
\nonumber \\
&- \Xloc^0(\iw; \iW)^2 V^4 
[ X_{\textrm{c},\bm{q}} (\iw; \iW) - \bar{X}_{\textrm{c}} (\iw; \iW) ].
\end{align}
Inserting this expression into $\Pq(\iw;\iW)$ in Eq.~(\ref{eq:P_q}) and retaining the term of order $V^4$, we obtain Eq.~(\ref{eq:Pq-PAM}).

\section{Technical details of SCL2 scheme}
\label{app:scl2}

\begin{figure}[tb]
	\begin{center}
	\includegraphics[scale=0.8]{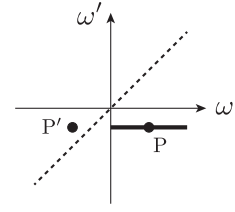}
	\end{center}
	\caption{A schematic figure explaining which data in $\Xloc(iw, iw')$ are used in the SCL2 scheme. The thick line is on $\omega'=\omega_0$, and points show $\textrm{P}=(\iw_1,-\iw_0)$ and $\textrm{P}'=(-\iw_1,-\iw_0)$.}
	\label{fig:scl2}
\end{figure}

Let us assume that the decoupling in Eq.~(\ref{eq:X-decouple}) is exactly satisfied, namely,
\begin{align}
\Xloc(\iw, \iw') &= s_0 u_0(\iw) v_0^*(\iw').
\end{align}
Then, $u_0(\iw)$ can be determined from a one-dimensional data of $\Xloc(\iw, \iw')$ as shown in Fig.~\ref{fig:scl_scheme}:
$u_0(\iw) \propto \Xloc(\iw, -\iw_0)$, where $\omega_0$ is, e.g., the lowest Matsubara frequency.
In practice, however, a good estimation is not achieved in this way, because actual data of $\Xloc(\iw, \iw')$ has a structure around the diagonal $\omega=\omega'$, which is not included in $u_0(\iw)$.
For better estimation of $u_0(\iw)$, we need to extract only the broad structure away from the diagonal.
For this reason, we use only the half data on $\omega>0$ side as depicted in Fig.~\ref{fig:scl2}, and the negative side is recovered using the symmetry relation in Eq.~(\ref{eq:u0-conj}).
In the following, we describe how to construct $u_0(\iw)$ in this way.

We first determine $\theta$, which is necessary to use Eq.~(\ref{eq:u0-conj}).
To this end, we input two data points, $\textrm{P}$ and $\textrm{P}'$, in Fig.~\ref{fig:scl2} and define
\begin{align}
\Theta &\equiv \arg [\Xloc(\iw_1,-\iw_0) + \Xloc(-\iw_1,-\iw_0)]
\nonumber \\
&= \frac{\theta}{2} + \arg [v_0^*(-\iw_0)].
\end{align}
Here, $\omega_1 \neq \omega_0$ is, e.g., the second lowest Matsubara point.
Then, we shift the phase of $\Xloc(\iw,-\iw_0)$ by $-\Theta$ to define
\begin{align}
\widetilde{u}_0(\iw) &\equiv u_0(\iw) e^{-i\theta/2}
\nonumber \\
&= C \Xloc(\iw,-\iw_0) e^{-i\Theta},
\end{align}
where $C=1/s_0 |v_0^*(-\iw)|$ is a real number.
The negative side, $\omega<0$, can be evaluated using the symmetry 
$\widetilde{u}_0(-\iw) = \widetilde{u}_0^*(\iw)$, which can be derived from Eq.~(\ref{eq:u0-conj}).
The factor $C$ is fixed from the normalization condition.
Finally, the singular value $s_0$ is determined using the relation $\Xloc(\iw_0,-\iw_0)=s_0 |u_0(\iw_0)|^2$.

\section{The function $u_0(i\omega)$ in a two-orbital model}
\label{app:scl3}

\begin{figure}[tb]
	\begin{center}
	\includegraphics[width=\linewidth]{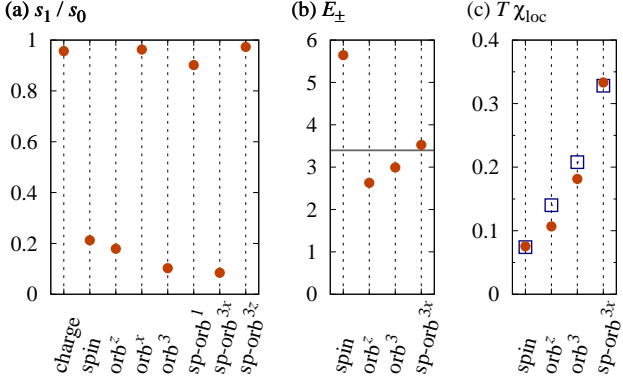}
	\end{center}
	\caption{The mode-dependences of several quantities in the two-orbital model: (a) The singular-value ratio $s^{\xi}_1/s^{\xi}_0$, (b) the fitting parameters, $E^{\xi}_+=E^{\xi}_-$, of the function $u_0^{\xi}(\iw)$ (the fitting results are depicted in Fig.~\ref{fig:u0-2orb}), and (c) $T\chiloc$ (circle) and $s_0 \overline{u}_0^2$ (square) for verification of of Eq.~(\ref{eq:s0-sumrule}). The horizontal line in (b) indicates the lowest excitation energy $E_{\pm, \textrm{min}}$.}
	\label{fig:params-2orb}
\end{figure}

\begin{figure}[tb]
	\begin{center}
	\includegraphics[width=\linewidth]{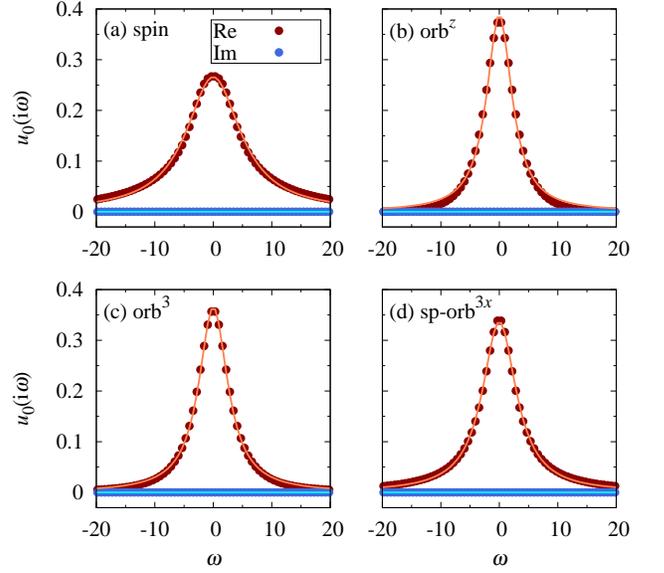}
	\end{center}
	\caption{The function $u_0^{\xi}(\iw)$ for the two-orbital model with $\Delta/W=1.1$ and $T=0.1$. The lines show fitting results by the function in Eq.~(\ref{eq:phi-gen}).}
	\label{fig:u0-2orb}
\end{figure}

In this Appendix, we present explicit results for the function $u_0^{\xi}(\iw)$ in the two-orbital model, and provide a grounding of the two-pole approximation introduced in Sec.~\ref{sec:SCL3}.

We first show the mode-dependence of the ratio $s^{\xi}_1/s^{\xi}_0$ in Fig.~\ref{fig:params-2orb}(a).
The data are taken from Fig.~\ref{fig:SV-multiorb} (the lowest temperature, $T=0.1$).
The decoupling approximation is justified if $s^{\xi}_1/s^{\xi}_0$ is small, and four modes, i.e., spin, orb$^{z}$, orb$^3$, and sp-orb$^{3x}$, apply to this case as expected.

The function $u_0^{\xi}(\iw)$ is shown for those four modes in Fig.~\ref{fig:u0-2orb}.
The lines show fitting results by the two-pole function in Eq.~(\ref{eq:phi-gen}), demonstrating that reasonable fitting was achieved.
The fitting parameter $E_+=E_-$ are depicted in Fig.~\ref{fig:params-2orb}(b).
The horizontal line indicates the lowest excitation energy $E_{\pm, \textrm{min}} \approx 3.40$ from $n=2$ to $n=2\pm1$.
The fitting results excepting the spin mode turn out to agree with $E_{\pm, \textrm{min}}$, indicating that replacing $E_{\pm}$ with $E_{\pm, \textrm{min}}$ provides a good approximation.
Regarding the spin mode, this replacement is expected to result in an overestimate of fluctuations.

Finally, we verify Eq.~(\ref{eq:s0-sumrule}), which relates $s_0^{\xi}$ to $\chiloc^{\xi}$.
Figure~\ref{fig:params-2orb}(c) compares the left- and right-hand sides of Eq.~(\ref{eq:s0-sumrule}), demonstrating that Eq.~(\ref{eq:s0-sumrule}) is satisfied in reasonable accuracy.
All the results above support the SCL3 approximation scheme as a simple and reasonable approximation for solving the BS equation.

\bibliography{JO,unpublished}

\end{document}